\documentclass[fleqn,usenatbib]{mnras}

\usepackage{newtxtext,newtxmath}
\usepackage{lscape}
\usepackage{rotfloat}

\usepackage[T1]{fontenc}

\DeclareRobustCommand{\VAN}[3]{#2}
\let\VANthebibliography\thebibliography
\def\thebibliography{\DeclareRobustCommand{\VAN}[3]{##3}\VANthebibliography}

\usepackage{graphicx}	
\usepackage{amsmath}	
\usepackage{siunitx} 
\usepackage{lscape}
\usepackage{xcolor}

\newcommand{\rev}[1]{#1}

\newcommand{\vh}{\mathbf{v_h}}
\newcommand{\divh}{\nabla_h \cdot \mathbf{v}_h}

\newcommand{\timthreetau}{$\tau=-0.8$~days } 
\newcommand{\tizerotau}{$\tau=-0.1$~days }
\newcommand{\ms}{~m~s$^{-1}$ }
\newcommand{\ps}{~s$^{-1}$ }

\title[A flux-independent increase in outflows prior to  active region emergence]{A flux-independent increase in outflows prior to the emergence of active regions on the Sun.}

\author[H.Schunker, et al.]{
H.~Schunker,$^{1}$\thanks{E-mail: hannah.schunker@newcastle.edu.au}
W.~Roland-Batty,$^{1}$
A.~C. Birch,$^{2}$
D.~C. Braun,$^{3}$
R.~H. Cameron,$^{2}$
L.~Gizon$^{2,4}$
\\
$^{1}$Centre of Solar and Space Physics, The University of Newcastle, Australia\\
$^{2}$Max-Planck-Institut f\"{u}r Sonnensystemforschung, 37077  G\"{o}ttingen, Germany\\
$^{3}$NorthWest Research Associates, 3380 Mitchell Lane, Boulder, CO 80301-2245, USA\\
$^{4}$Institut f\"{u}r Astrophysik und Geophysik, Georg-August-Universit\"{a}t G\"{o}ttingen, 37077 G\"{o}ttingen, Germany
}

\date{Accepted XXX. Received YYY; in original form ZZZ}

\pubyear{2024}

\begin{document}
\label{firstpage}
\pagerange{\pageref{firstpage}--\pageref{lastpage}}
\maketitle

\begin{abstract}
\rev{Emerging active regions are associated with convective flows on the spatial scale and lifetimes of supergranules. To understand how these flows are involved in the formation of active regions, we aim to identify where active regions emerge in the supergranulation flow pattern. We computed supergranulation scale flow maps at the surface for all active regions in the Solar Dynamics Observatory Helioseismic Emerging Active Region Survey. We classified each of the active regions into four bins, based on the amplitude of their average surface flow divergence at emergence. We then averaged the flow divergence over the active regions in each bin as a function of time. We also considered a corresponding set of control regions. We found that, on average, the flow divergence increases during the day prior to emergence at a rate independent of the amount of flux that emerges. By subtracting the averaged flow divergence of the control regions, we found that active region emergence is associated with a remaining converging flow at 0.5-1~days prior to emergence. This remnant flow, $\Delta \, \mathrm{div} \, \mathbf{v_h}  = (-4.9 \pm 1.7) \times 10^{-6}$\ps, corresponds to a flow speed of 10-20\ms (an order of magnitude less than supergranulation flows) out to a radius of about 10~Mm. 
We show that these observational results are qualitatively supported by simulations of a small bipole emerging through the near-surface convective layers of the Sun. The question remains whether these flows are driving the emergence, or are caused by the emergence.}
\end{abstract}




\section{Introduction}

Active regions on the Sun are assumed to be a direct consequence of the Sun's global magnetic dynamo, and are thought to be caused by magnetic flux concentrations rising in a loop-like structure through the surface of the Sun \citep[e.g.][]{SpiegelWeiss1980,Cameronetal2017}. Alternatively, these magnetic flux concentrations could form in the near-surface layers \citep{Brandenburg2005} or throughout the convection zone \citep[e.g.][]{Nelson2014}. Understanding how active regions form will place constraints on their origins, and thus the location of the solar dynamo in the Sun's interior \citep[e.g. see][for a recent discussion]{Weberetal2023}.

Local helioseismology, a technique to probe the subsurface structure and flows of the Sun in three dimensions \citep{GizonBirch2005}, offers the potential to image the subsurface prior to an active region appearing at the surface.
Numerous case studies applying local helioseismology to emerging active regions \citep[e.g.][]{Kommetal2008,Kommetal2009,Ilonidisetal2011} did not result in a consensus view of the subsurface changes, if any, associated with the pre-emergence stage of active region formation. In addition, any surface magnetic field may contaminate the helioseismic signal \citep[e.g.][]{Braunetal1987,Schunkeretal2008,Schunkeretal2013}. 
 A plausible reason for this ambiguity is that realisation noise and background convection, dominated by supergranulation, can mask  weaker emergence signatures \citep{Birchetal2010}. This has lead to a need for improved statistics provided by examining many emerging regions.

\cite{Birchetal2013} used helioseismic holography \citep{LindseyBraun2000} to measure the subsurface flows prior to the formation of one hundred active regions observed by the Global Oscillation Network Group \citep[GONG,][]{Harveyetal1998}. They found that there were no statistically significant flows below the surface, however near the surface they found a statistically significant flow of about 15~\ms towards the emergence location in the day preceding the active region formation.

Extending the approach of \cite{Birchetal2013}, \cite{Schunkeretal2016} and \cite{Schunkeretal2019} identified a sample of 182  emerging active regions observed by the Helioseismic and Magnetic Imager on board the Solar Dynamics Observatory (SDO/HMI) \citep{SDO2012}.  The SDO/HMI observations are higher resolution than GONG and the Michelson Doppler Imager (MDI) allowing reliable helioseismic measurements to be made closer to the limb, and thus further back in time from the emergence, and they also have a significantly higher duty cycle. This data set is called the SDO Helioseismic Emerging Active Regions Survey (SDO/HEARS).

\cite{Birchetal2016} measured the flows, using both helioseismic holography and local correlation tracking, at the emergence of the SDO/HEARS active regions. They then compared the surface flows with those of simulations of a flux tube rising at different speeds, in a similar way to \cite{RempelCheung2014}, and showed that flux tubes rising faster than about 100\ms produced diverging flows at the surface which were not observed as active regions formed. Following this, \cite{Birchetal2019} averaged the surface flows over all active regions showing an east-west elongated, converging flow of about 40\ms, with the active region emergence located at the prograde end. They also showed that this flow pattern can be largely reproduced by a simple model in which active region emergence occurs preferentially in the prograde direction relative to supergranulation inflows, suggesting that rising flux concentrations and supergranule-scale flows interact during the emergence process. These results, in addition to others such as \cite{Schunkeretal2016,Schunkeretal2019,Schunkeretal2020}, point to a \emph{passive} emergence process driven largely by the convective flows at supergranulation scales, \rev{as opposed to an \emph{active} emergence where the characteristics of the emergence depends on magnetic flux.}

Motivated by the  results of \cite{Barnesetal2014} who showed that the single best indicator of an imminent active region emergence is the magnitude of the surface magnetic flux,  \cite{AlleySchunker2023} identified 42 emerging active regions in the SDO/HEARS which showed persistent bipoles more than two days before emergence, and 42 which clearly did \emph{not} have any pre-emergence bipoles. They found that these two samples had  distinct average surface flow divergence patterns. Averaging the flow maps of active regions with pre-emergence bipoles showed a statistically significant converging flow of about 100\ms, followed by no significant flow post-emergence. However, averaging over the flow maps of active regions that appeared  abruptly showed flow divergence of about 100\ms post emergence, but no significant flow prior to emergence. In addition, they found that the active regions with pre-emergence bipoles developed only into weak, low flux active regions, whereas the abruptly emerging active regions developed with higher flux.

Here we reconcile the results of \cite{Birchetal2019} and \cite{AlleySchunker2023} to define the location of active regions in the supergranulation pattern more closely. Our goal is to determine where active regions emerge in relation to the supergranular convection pattern, and if there is a dependence on the magnetic flux. We expect that these flows will be useful for constraining models for the origin and formation of active regions.

\section{Observations of Emerging Active Regions}

The Solar Dynamics Observatory and Helioseismic Emerging Active Regions Survey \citep[SDO/HEARS][]{Schunkeretal2016} consists of 182 emerging active regions observed by SDO/HMI between May 2010 and July 2014. \rev{To account for any systematics, each emerging active region was paired with a quiet-Sun control region tracked at the same disk position. The second requirement for these regions is that an active region does not emerge within the central $20^\circ$ radius. For all cases the control regions are mostly within ten days of the active region emergence.}

Each full-disk SDO Dopplergram, magnetogram and intensity continuum image was remapped using a Postel projection, centred on the active region \citep[see Table A.1 of][]{Schunkeretal2016}. Each resulting map has a pixel size of 1.39~Mm and contains $512 \times 512$ grid points. The $x$ and $y$ coordinates in the remapped images have $x$ increasing in the westward direction (prograde direction) and $y$ increasing in the northward direction. The location of the active regions was tracked at the Carrington rotation rate up to seven days before and after the emergence. The emergence time of each active region is based on the value of the magnetic flux measured with a 12 minute cadences \citep[see ][ for full details]{Schunkeretal2016}.

For helioseismology purposes the data is divided into 6.825-hour-long datacubes (547 frames), and are labelled with a time interval (\texttt{TI}) relative to the emergence time interval ( \texttt{TI+00}). Table~B.1 in \citet{Schunkeretal2019} lists the mid-time of the averaged \texttt{TI} to the time of emergence, $\tau=0$, for each time interval label. In this manuscript,  particularly relevant are the time intervals \texttt{TI-03}, equivalent to \timthreetau, and \texttt{TI+00}, equivalent to \tizerotau.

Not all active regions have observations at each time interval, depending on where they emerged on the solar disk and the duty cycle of the SDO full-disk observations. For example, at \texttt{TI-03} there are flow and magnetic field maps for 172 emerging active regions (EARs), at \texttt{TI+00} there are maps for 174 EARs, and at \texttt{TI+05} 177 EARs. In principle, we have a datacube for intensity, velocity and line-of-sight magnetic field for each EAR at each time interval that it crossed the disk.

\subsection{Computing the surface flows}

We computed the surface flows from the Doppler velocity datacubes at each time interval using surface-focusing holography as described in \cite{Birchetal2016}. We filtered the remapped Doppler velocities with a phase-speed filter with a central phase speed of $17.49$~km s$^{-1}$ and a width of $2.63$~km s$^{-1}$ \citep[filter 3 from Table~1 from the work of ][]{Couvidatetal2005}, selecting waves sensitive to the top 3~Mm below the surface. We used helioseismic holography \citep{LindseyBraun2000} focused at the surface to measure the north-south and east-west travel time differences. We then used an empirically determined conversion constant of $-7.7$~m~s$^{-2}$ to calibrate the east-west and north-south travel-time differences  to flows in units of \ms.  
\rev{We subtracted the best-fit second-order polynomial in two dimensions, which captures the large-scale effects on the surface of a sphere (such as differential rotation and meridional flow), from each map \rev{as is good practice \citep[e.g.][]{Birch+2013}}.}
We refer to these calibrated travel-time maps as $v_x$ (westward flow, positive for prograde flows) and $v_y$ (northward flow, positive for poleward flows).

After computing the flow maps, we applied a filter to reduce the contribution of realisation noise and any remnant large scale flows or systematics to the travel-time maps. The filter had a value of one for angular degree $90 < kR_\odot < 140$, tapering with a raised cosine to zero for $kR_\odot < 10$ and $kR_\odot > 220$, \rev{where $k$ is the spatial wavenumber in Fourier space, retaining flows on supergranulation length scales \cite[$\approx 20-40$~Mm, e.g.][]{Gizon+2003}}.
The focus in the current work is the local supergranulation scale flows associated with the emergence process. Supergranules can be most easily identified in maps of the flow divergence $\mathrm{div} \, \mathbf{v}_h = \divh$, where the horizontal vector velocity $\vh$  is given by ($v_x$,$v_y$). \rev{We emphasise that negative (positive) flow divergence indicates a converging (diverging) flow.}

\section{Classification of surface flow divergence}\label{sect:classflows}

\begin{figure}
	\includegraphics[width=0.5\textwidth]{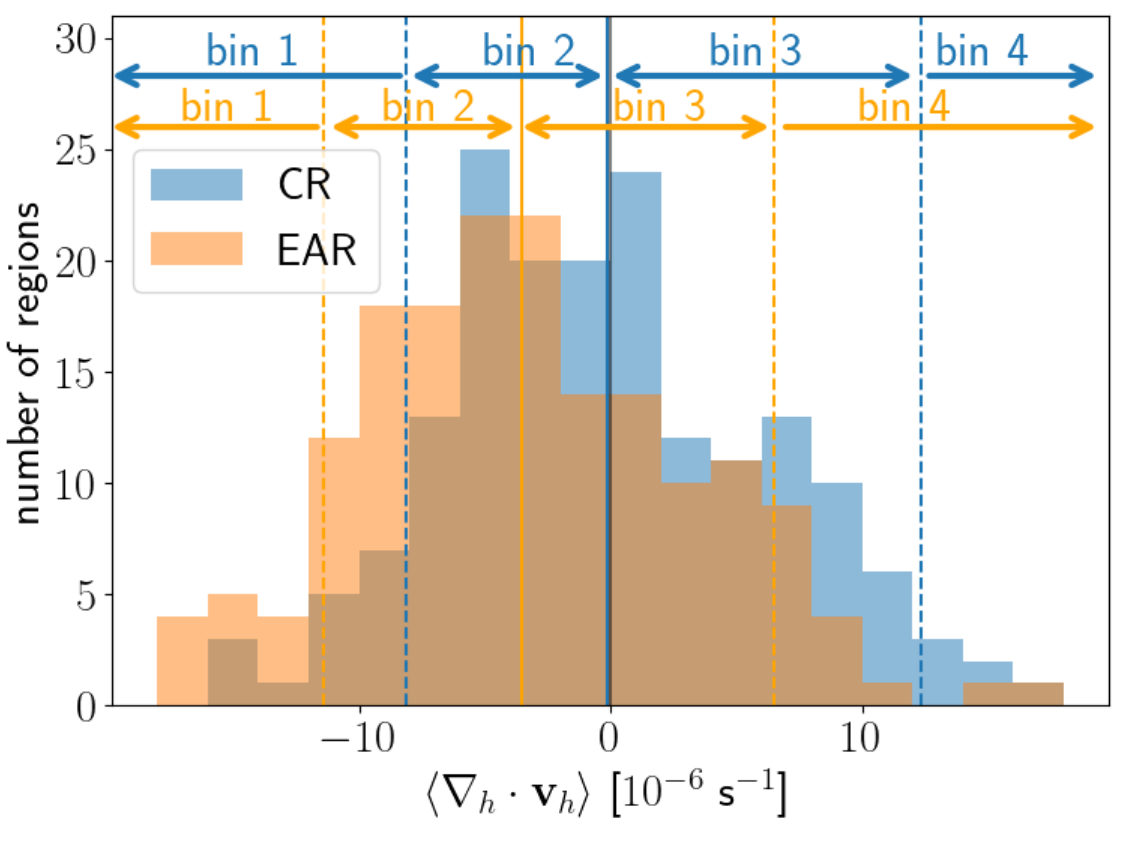}
    \caption{Distribution of flow divergence averaged over the central 10~Mm radius of the maps at \timthreetau  (\texttt{TI-03}) for the 172 control regions (blue) and 172 EARs (orange) with divergence and magnetic field maps. The vertical blue lines indicate the values used to bin the flows in the control sample at values 
    $\langle \divh \rangle_r = -8.1 \times 10^{-6}, 0.0 \times 10^{-6} \, \textrm{(solid blue line)}, \, \textrm{and} \, 12.3 \times 10^{-6}$~\ps. 
    The vertical orange  lines indicate the values used to bin the flows of EARs at values: 
    $\langle \divh \rangle_r = -11.4 \times 10^{-6},  -3.5 \times 10^{-6} \, \textrm{(solid orange line)}, \, \textrm{and} \, 6.5 \times 10^{-6}$~\ps. Note that the end bins of the histograms contain all values beyond the last interval, some of which lie beyond the plot range.}
    \label{fig:voihistom3}
\end{figure}

\cite{Birchetal2019} found that active regions prefer to emerge in converging flows at \timthreetau  (time interval \texttt{TI-03}).
To classify these flows, we average the flow divergence in the central 10~Mm radius of each map, $\langle \divh \rangle_r$, for all active and control regions at  \timthreetau. We selected a radius of 10~Mm  by inspection to isolate the central flow divergence signature in the map. Fig.~\ref{fig:voihistom3} shows a histogram of the flows for all emerging active regions (EARs) and control regions (CRs) at this time interval.  The mean of the EAR flows is indeed a converging flow (solid vertical orange line), consistent with \cite{Birchetal2016}. The mean of the CR flows is zero. We classify the flow distribution into four bins, defined first by the mean of the full sample, and then each half is separated by the mean of the end points of the range. The boundaries of the control region bins are: 
\begin{eqnarray*}
 \langle \divh \rangle_{r,\mathrm{CR}} - \left( \frac{ \langle \divh \rangle_{r,\mathrm{CR}} - \mathrm{min}( \langle \divh \rangle_r )}{2}\right)& = -8.1~\times 10^{-6}\\
 \langle \divh \rangle_{r,\mathrm{CR}}& = 0.0 \times 10^{-6} \\
\langle \divh \rangle_{r,\mathrm{CR}} + \left(\frac{ \mathrm{max} ( \langle \divh \rangle_r ) - \langle \divh \rangle_{r,\mathrm{CR}}}{2}\right)& = 12.3 \times 10^{-6}
\end{eqnarray*}
in units of \ps (see the blue vertical lines in Fig.~\ref{fig:voihistom3}).  We  used the subscript $r$ and $\mathrm{CR}$ to clarify when the average is over radius only, or radius and over all control regions.  The minimum average divergence value is $\mathrm{min}\left( \langle \divh \rangle \right) = -16.2 \times 10^{-6}$~\ps \, (from the control region for AR~11206) and the maximum average divergence value is $\mathrm{max} \left( \langle \divh \rangle \right) = 24.8 \times 10^{-6}$~\ps \, (from the control region for AR~11074). Presumably, the first bin (CR bin 1, with the strongest flow convergence) selects inflows between supergranules, and the last bin (CR bin 4, with the strongest flow divergence) selects for the centres of well-developed supergranules.

Equivalently for the EARS, the boundaries of the bins  are:
\begin{eqnarray*}
\langle \divh \rangle_{r,\mathrm{EAR}} - \left( \frac{\langle \divh \rangle_{r,\rev{\mathrm{EAR}}} - \textrm{min}(\langle \divh \rangle_\mathrm{r})}{2}\right) &= -11.4 \times 10^{-6},\\
\langle \divh \rangle_{r,\mathrm{EAR}} &= -3.5 \times 10^{-6}, \\
\mathrm{and} \hspace{4.8cm} & \\
\langle \divh \rangle_{r,\mathrm{EAR}} + \left(\frac{ \textrm{max}(\langle \divh \rangle_r) - \langle \divh \rangle_{r,\mathrm{EAR}}}{2}\right)&= 6.5 \times 10^{-6}
\end{eqnarray*}
in units of \ps (see orange vertical lines in Fig.~\ref{fig:voihistom3}). The minimum average divergence value is $\mathrm{min} \left( \langle \divh \rangle_r \right) = (-1.9  \pm 0.6)  \times 10^{-6}$~\ps \, (from AR~11194) and the maximum average divergence value is $\mathrm{max} \left( \langle \divh \rangle_r \right) = (16.6 \pm 1.1) \times 10^{-6}$~\ps \, (from AR~11706). 

 \section{Averaged flow divergence and magnetic field maps}\label{sect:avemaps}

We averaged the line-of-sight magnetic field datacubes at each time interval so that we had one averaged magnetic field map corresponding to each flow divergence map. Under the assumption that the magnetic field is mostly radial at the solar surface, we approximately corrected for the line-of-sight projection of the magnetic field by dividing it by $\cos \Theta$, where $\Theta$ is the angular distance to disk centre \citep{Leka+2017}.
We then shifted the averaged line-of-sight magnetogram and flow divergence map, using bilinear interpolation over the four nearest pixels, so that the centre of the map coincides with the centre of the active region  \citep[as defined by][]{Birchetal2013}. 

\begin{figure*}
	\includegraphics[width=0.9\textwidth]{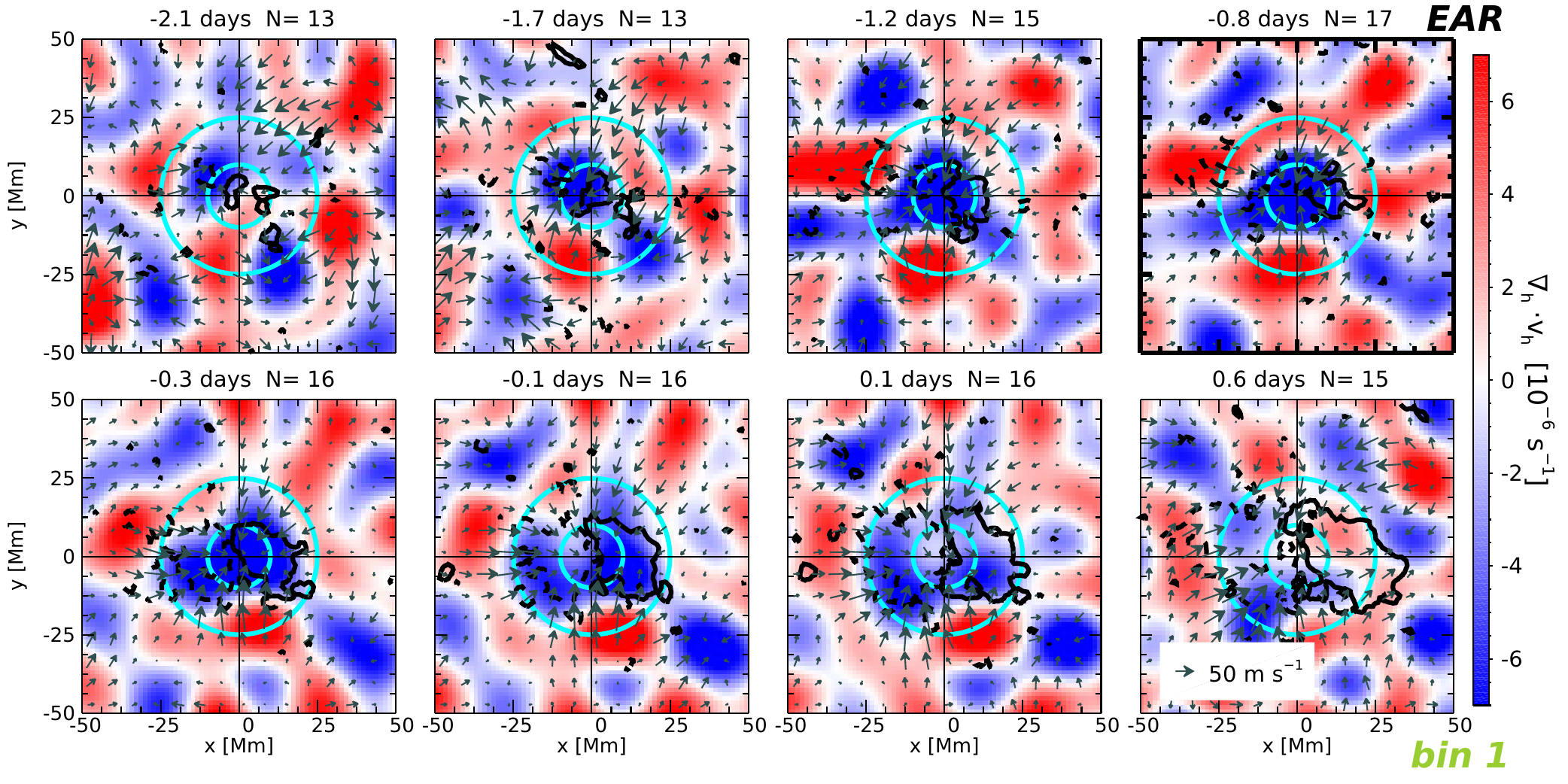}
    \caption{Averaged divergence flow maps for active regions with  $\langle \divh \rangle_r \leq -11.4 \times 10^{-6}$~\ps (EAR, bin 1 \rev{with the strongest converging flows at $\tau=-0.8$~days of the entire sample}). The arrows indicate the magnitude and direction of the horizontal flows. The solid black contour represents $+20$~G and the dashed contour $-20$~G. The number of maps contributing to the average is the number $N$. The inner cyan circle outlines the area within which the flow divergence is averaged and the outer circle outlines the area within which the absolute magnetic field is averaged. The thick axis represents the time interval (\texttt{TI-03}) that was used to classify the active regions. The maps correspond to every second time intervals \texttt{TI-09, TI-07, TI-05, TI-03, TI-01, TI+00, TI+01, TI+03}. }
    \label{fig:avedivmapso1m3}
\end{figure*}

\begin{figure*}
	\includegraphics[width=0.9\textwidth]{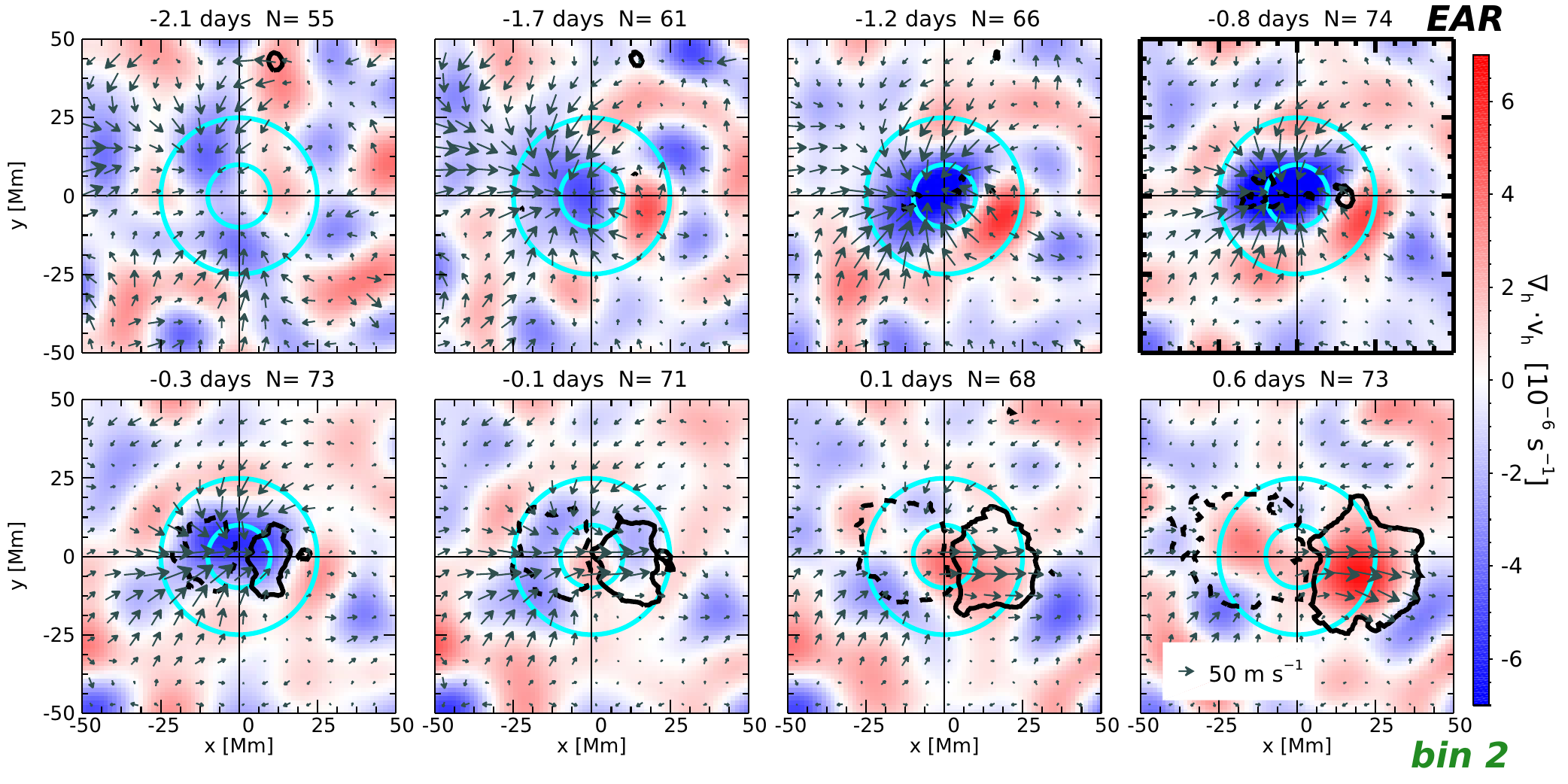}
    \caption{Averaged divergence flow maps for active regions  with  $ -11.4 \times 10^{-6} < \langle \divh \rangle_r \leq -3.5 \times 10^{-6}$~\ps (EAR bin 2). \rev{All other annotations are the same as Fig.~\ref{fig:avedivmapso1m3}}.}
    \label{fig:avedivmapso2m3}
\end{figure*}

\begin{figure*}
	\includegraphics[width=0.9\textwidth]{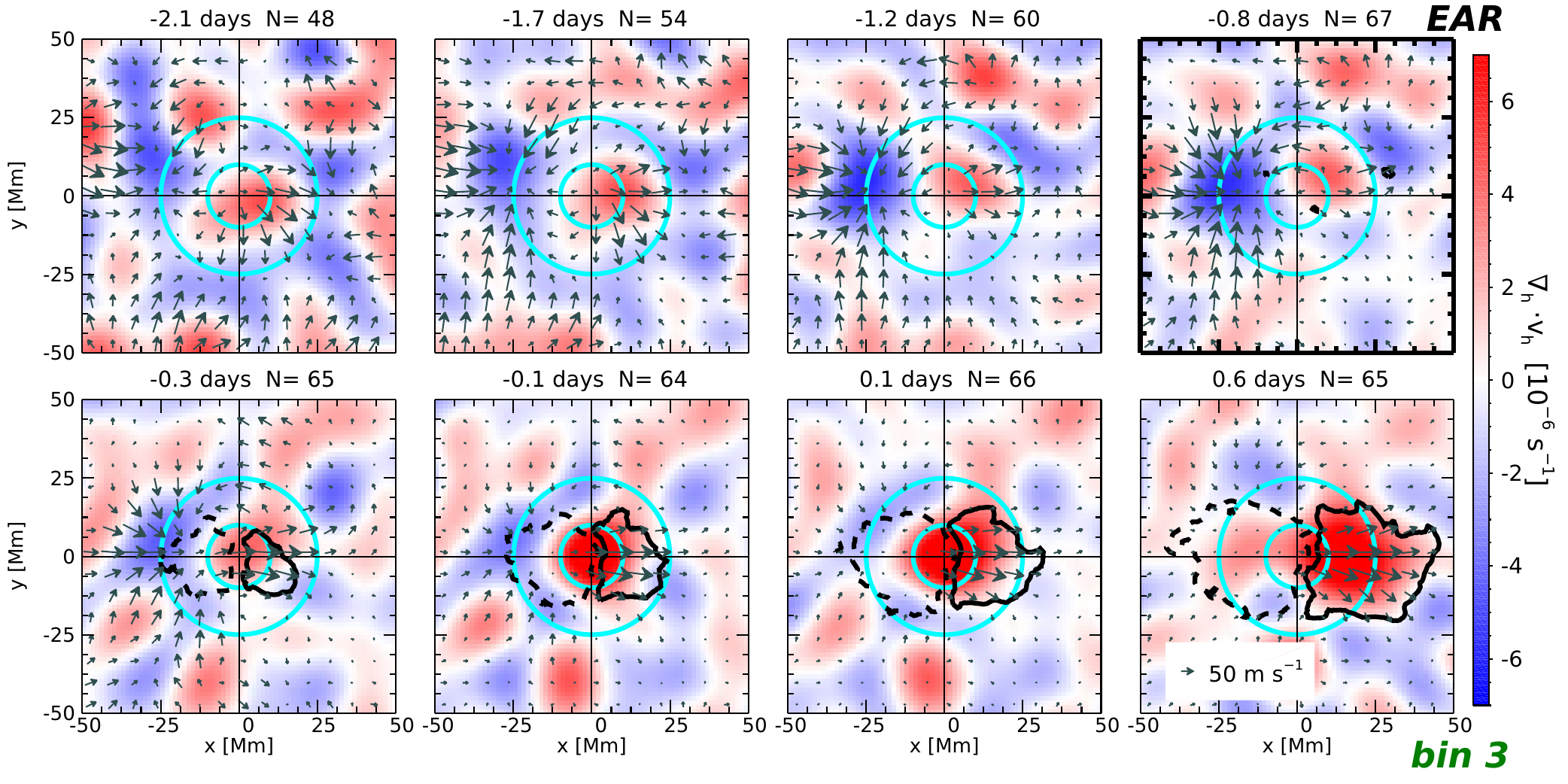}
    \caption{Averaged divergence flow maps for active regions  with  $ -3.5 \times 10^{-6} < \langle \divh \rangle_r \leq 6.5 \times 10^{-6}$~\ps (EAR, bin 3). \rev{All other annotations are the same as Fig.~\ref{fig:avedivmapso1m3}}.}
    \label{fig:avedivmapso3m3}
\end{figure*}

\begin{figure*}
	\includegraphics[width=0.9\textwidth]{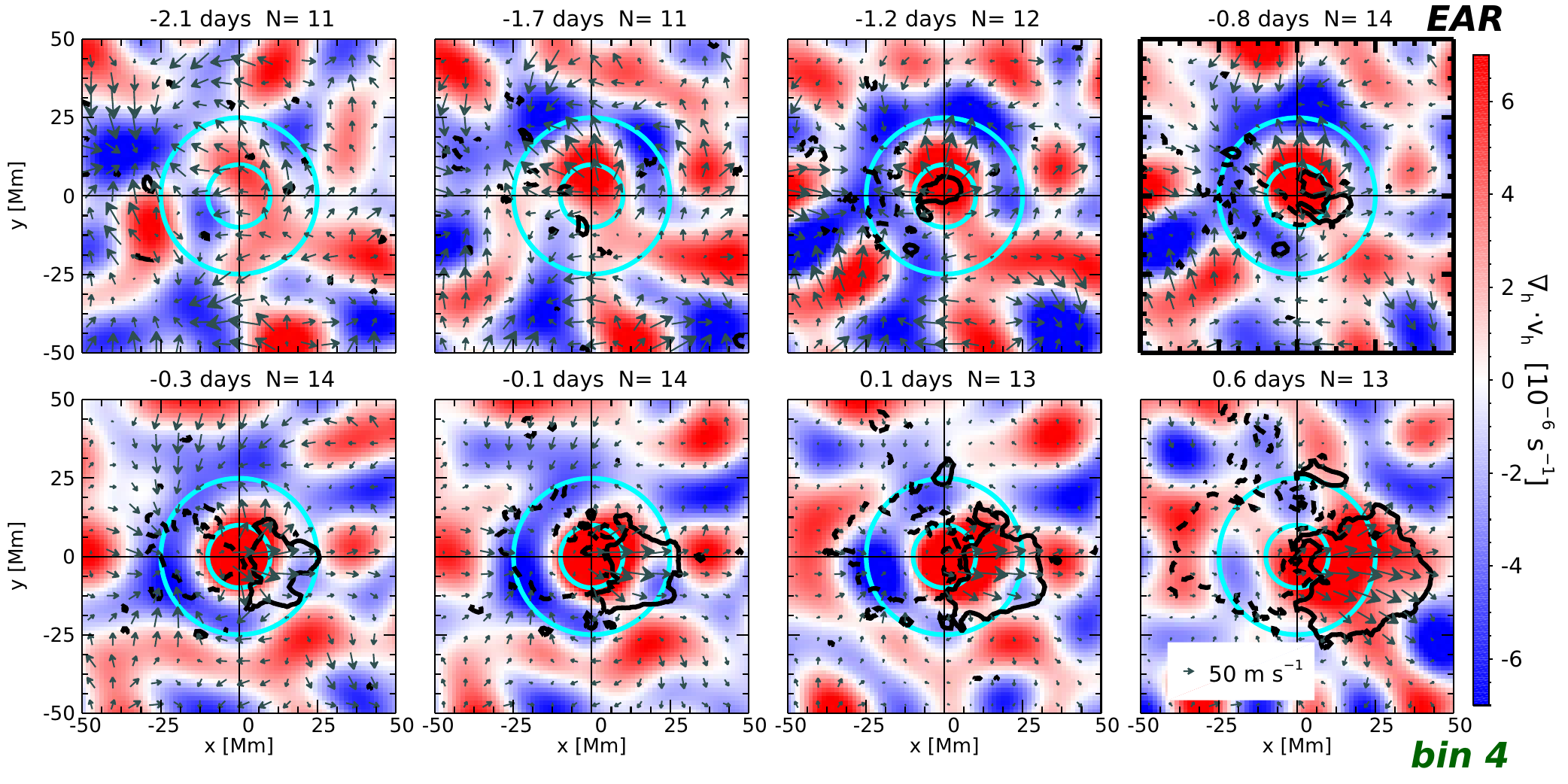}
    \caption{Averaged divergence flow maps for active regions with $\langle \divh \rangle_r > 6.5 \times 10^{-6}$~\ps (EAR, bin 4 \rev{with the strongest diverging flows at $\tau=-0.8$~days of the entire sample}). \rev{All other annotations are the same as Fig.~\ref{fig:avedivmapso1m3}}.}
    \label{fig:avedivmapso4m3}
\end{figure*}

\begin{figure*}
	\includegraphics[width=0.9\textwidth]{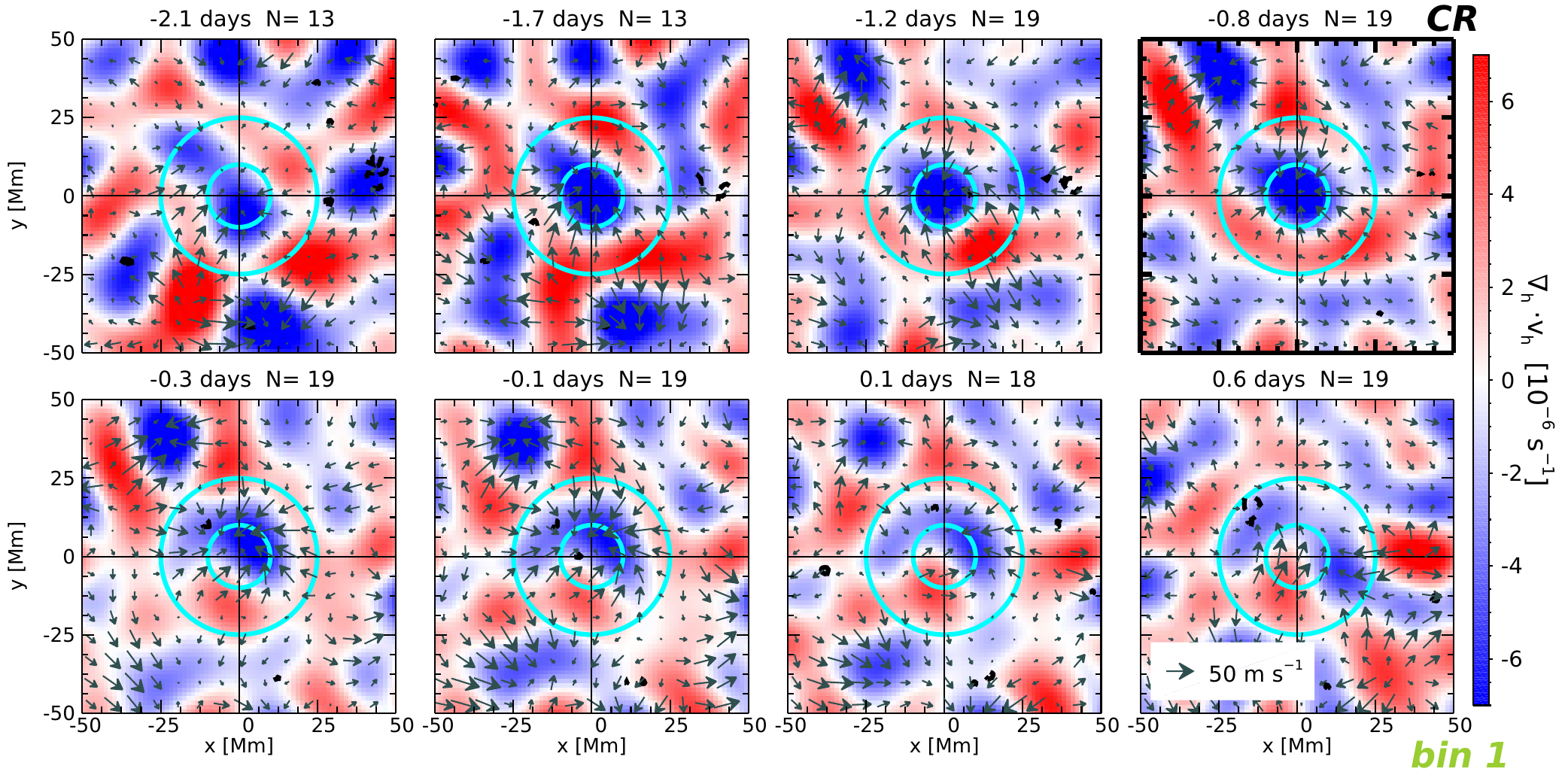}
    \caption{Averaged divergence flow maps for control regions at \texttt{TI-03} or \timthreetau with  $\langle \divh \rangle_r < -8.1 \times 10^{-6}$~\ps (CR bin 1, \rev{with the strongest converging flows at $\tau=-0.8$~days from all of the control regions}). All other annotations are the same as Fig.~\ref{fig:avedivmapso1m3}.}
    \label{fig:avedivmapsqo1}
\end{figure*}

\begin{figure*}
	\includegraphics[width=0.9\textwidth]{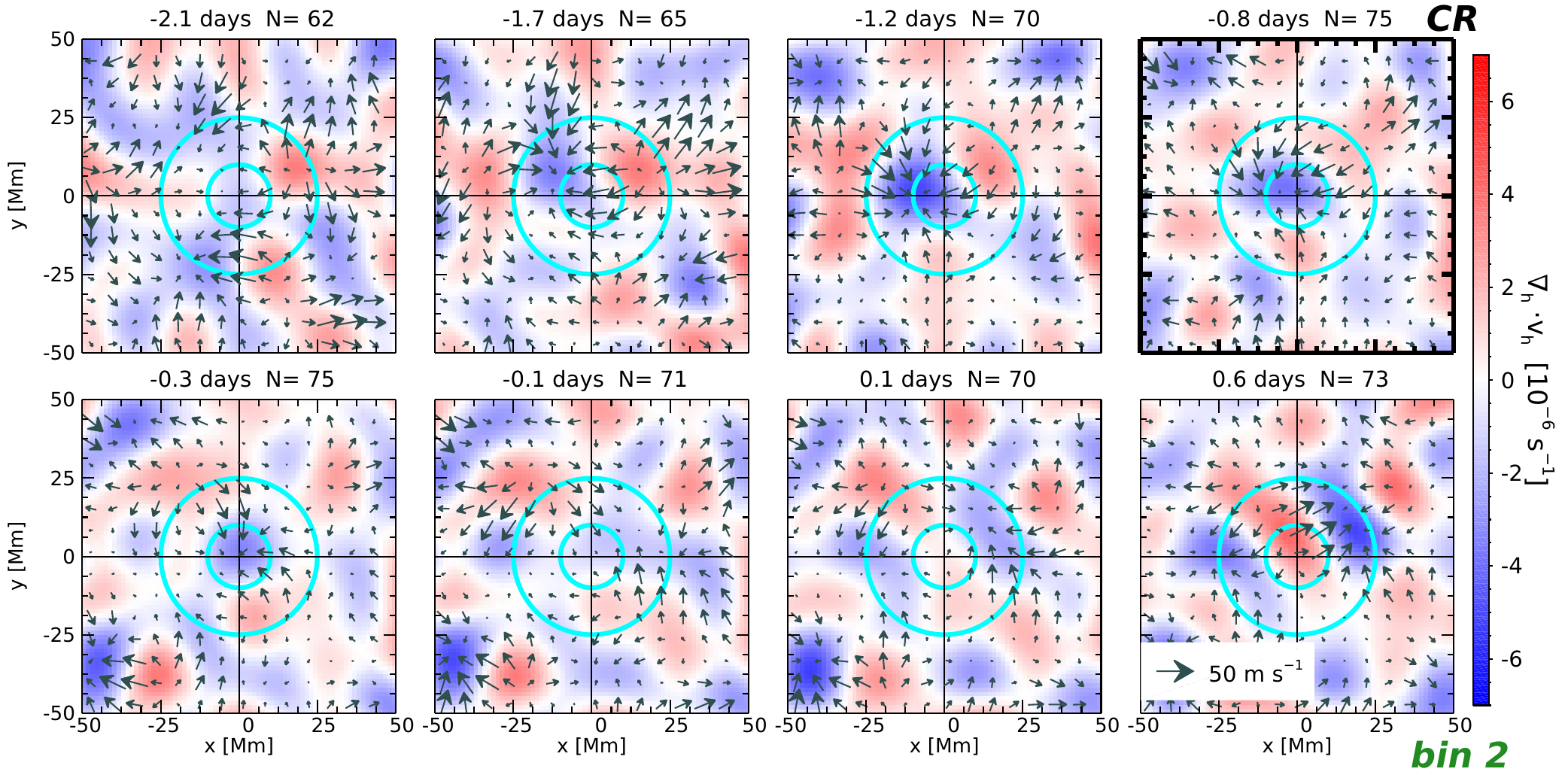}
    \caption{Averaged divergence flow maps for control regions at \texttt{TI-03} or \timthreetau  with  $-8.1 \times 10^{-6} < \langle \divh \rangle_r < 0.0 \times 10^{-6}$~\ps (CR, bin 2). The arrows indicate the magnitude and direction of the horizontal flows. \rev{All other annotations are the same as Fig.~\ref{fig:avedivmapso1m3}}. }
    \label{fig:avedivmapsqo2}
\end{figure*}

\begin{figure*}
	\includegraphics[width=0.9\textwidth]{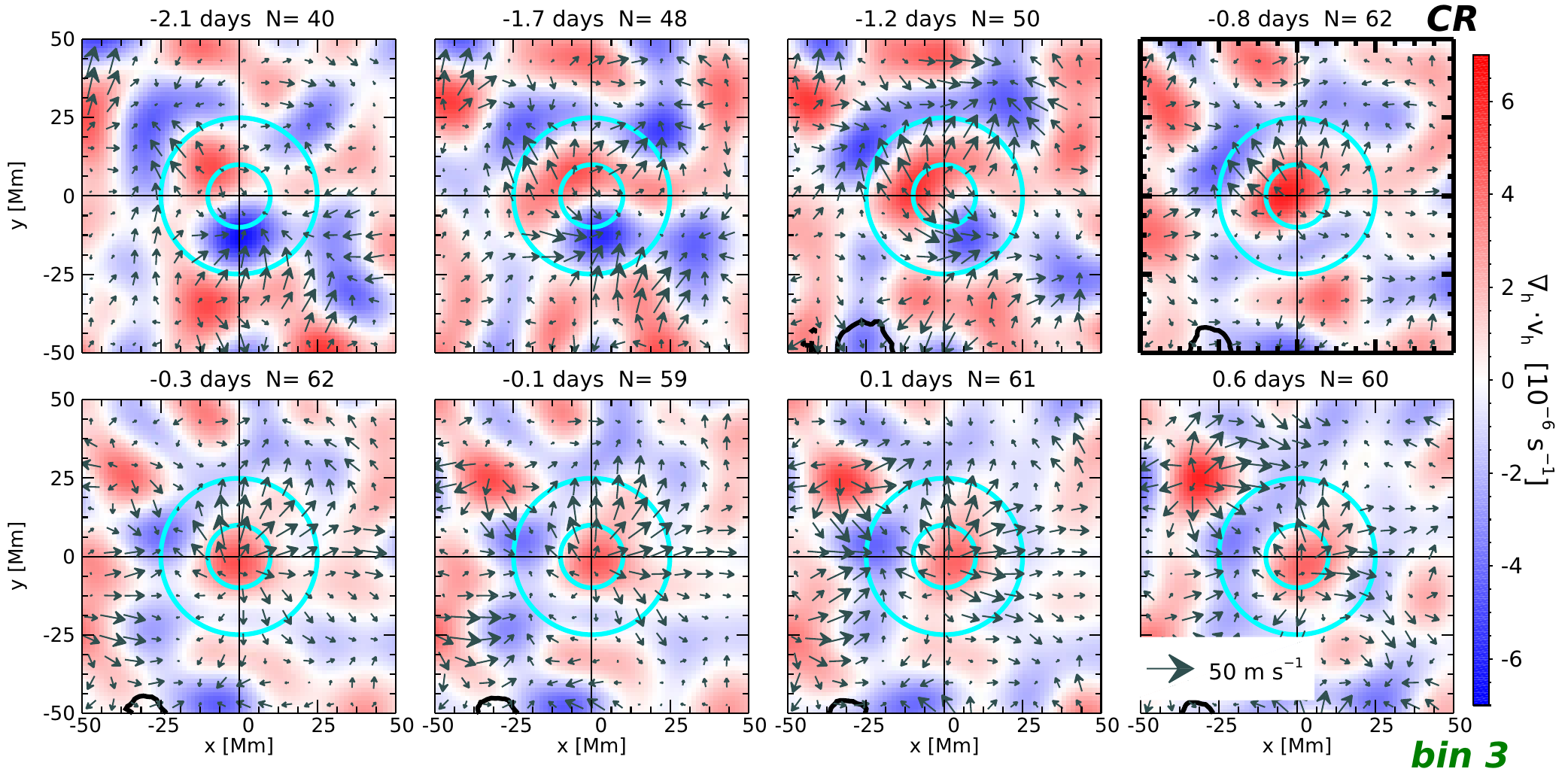}
    \caption{Averaged divergence flow maps for control regions  at \texttt{TI-03} or \timthreetau with  $0.0 \times 10^{-6} < \langle \divh \rangle_r < 12.3 \times 10^{-6}$~\ps (CR, bin 3). \rev{All other annotations are the same as Fig.~\ref{fig:avedivmapso1m3}}.}
    \label{fig:avedivmapsqo3}
\end{figure*}

\begin{figure*}
	\includegraphics[width=0.9\textwidth]{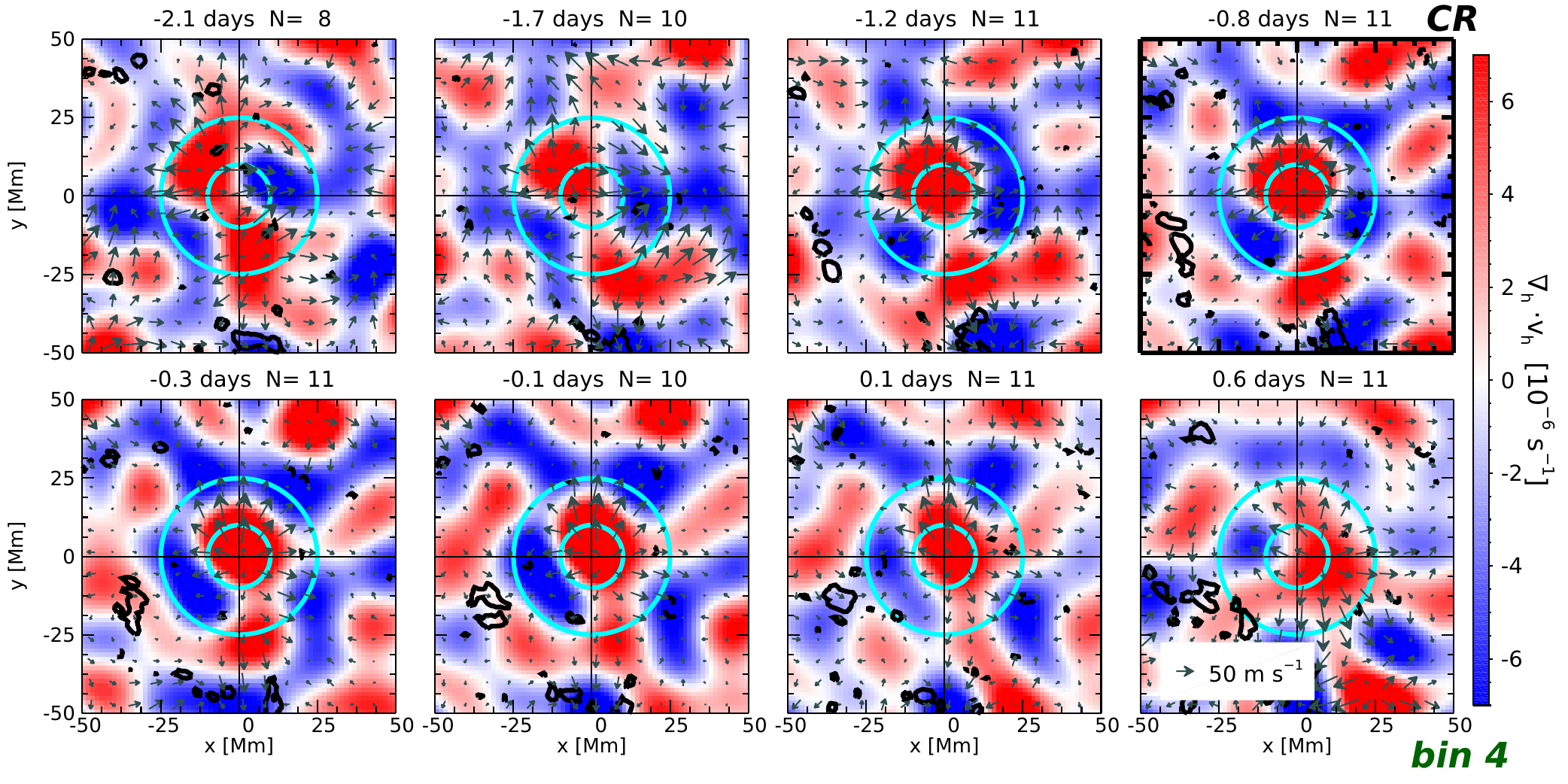}
    \caption{Averaged divergence flow maps for control regions at \texttt{TI-03} or \timthreetau with  $ \langle \divh \rangle_r > 12.3 \times 10^{-6}$~\ps (CR, bin 4 \rev{with the strongest diverging flows at $\tau=-0.8$~days from all of the control regions}). \rev{All other annotations are the same as Fig.~\ref{fig:avedivmapso1m3}}.}
    \label{fig:avedivmapsqo4}
\end{figure*}

 Figures~\ref{fig:avedivmapso1m3} - \ref{fig:avedivmapso4m3} show the averaged flow divergence maps in each bin as a function of time. Figure~\ref{fig:avedivmapso1m3} shows the evolution of the flows for the sample with the strongest converging flows at \timthreetau (\texttt{TI-03}). The strong converging flows at the emergence location are also significant leading up to this time interval, however, beyond, it becomes weaker. Fig.~\ref{fig:avedivmapso4m3}, in contrast, shows the evolution of the strongest flow divergence at \timthreetau which remains significant for future times. These maps in Fig.~\ref{fig:avedivmapso1m3} and Fig.~\ref{fig:avedivmapso4m3} are noisier given that they are only averaged over about fifteen EARs, whereas Figs.~\ref{fig:avedivmapso2m3} and \ref{fig:avedivmapso3m3} are averaged over two to three times that many maps.  
Each of these time series shows there are persistent and significant flows at the site of the active region emergence, in comparison to the surrounding areas of the averaged flow divergence maps and those for the control regions (Figures~\ref{fig:avedivmapsqo1}-~\ref{fig:avedivmapsqo4}).

The central flows in the control regions with the largest converging flows  at $\tau=-0.8$~days (shown in Fig.~\ref{fig:avedivmapsqo1}) become insignificant compared to the surrounding flows at  $\tau=0.6$~days, reflecting the supergranulation lifetime of 1-2~days. This is also true for the central flow regions with the largest diverging flow  (Fig.~\ref{fig:avedivmapsqo4}), indicating that our control regions are selecting supergranules at various stages of their lifetime.

\begin{figure*}
	\includegraphics[width=1.0\textwidth]{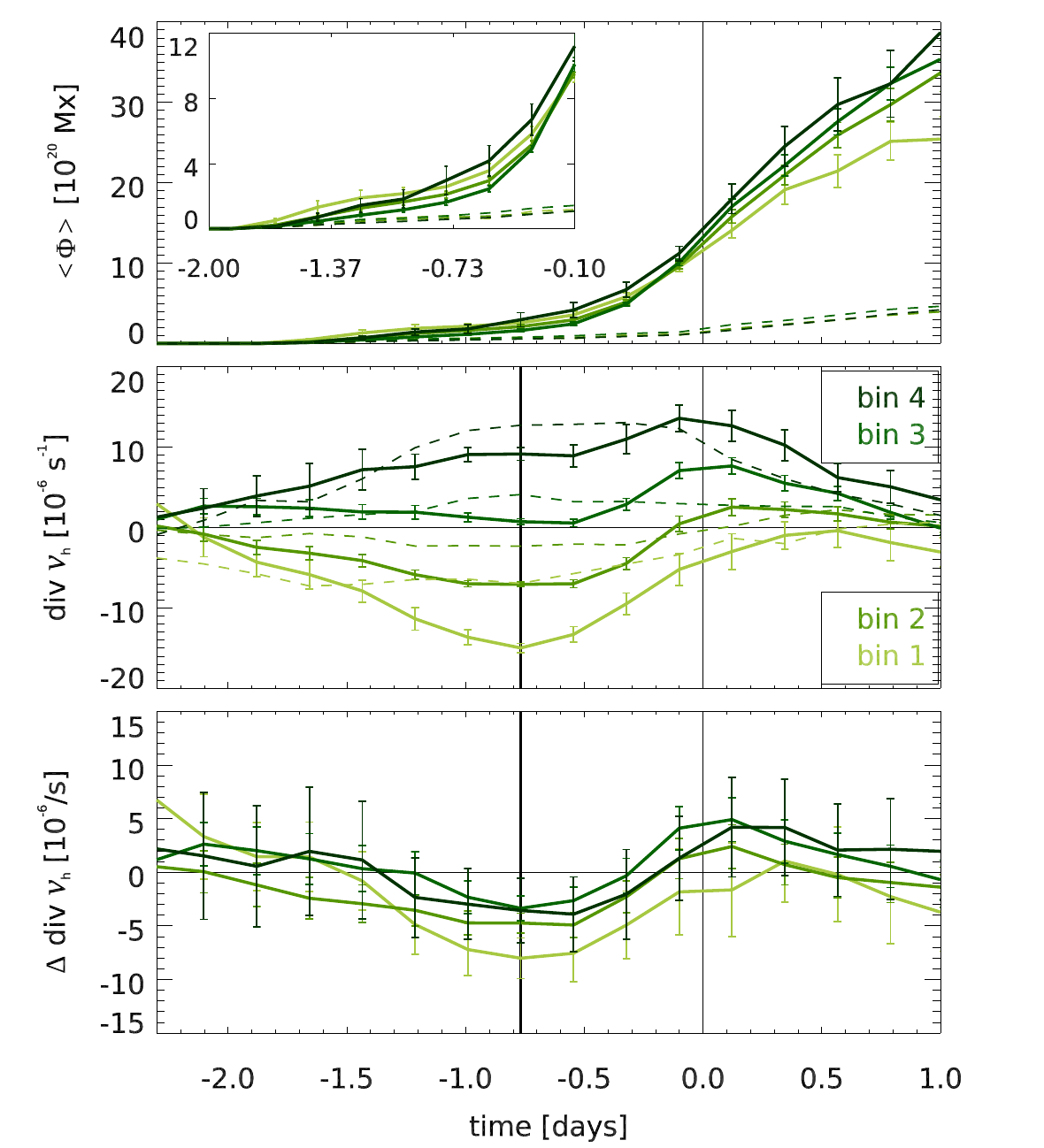}
 \vspace{0cm}
    \caption{Mean magnetic flux and divergence flows during the emergence process. The EARS were divided into four bins (see Fig.~\ref{fig:voihistom3}) based on the value of the mean divergence in the central 10~Mm radius at \timthreetau (\texttt{TI-03}) indicated by the colour gradient of the curves, from the strongest convergence regions in light green (bin 1) through to the strongest divergence regions in dark green (bin 4).  The thick vertical black line indicates the time at which the bins were defined. The top panel shows the mean flux, $\langle \Phi \rangle_{r,\mathrm{EAR}}$ (solid curve) and $\langle \Phi \rangle_{r,\mathrm{CR}}$ (dashed curve), in the central 25~Mm radius.   The middle panel shows the mean divergence, $\mathrm{div} v_h = \langle \divh \rangle_{r,\mathrm{EAR}}$ (solid curve) and $\mathrm{div} v_h = \langle \divh \rangle_{r,\mathrm{CR}}$ (dashed curve), of the surface flows.  The bottom panel shows the \rev{difference, $\Delta \, \mathrm{div} \, \mathbf{v}_h = \langle \divh \rangle_{r,\mathrm{EAR}} - \langle \divh \rangle_{r,\mathrm{CR}}$}, in average flow divergence between the control regions and active regions from the middle panel. \rev{The uncertainties are the standard error over all active regions in each bin at each time interval.}}
    \label{fig:avedivflowsm3}
\end{figure*}

We compared the averaged surface flows in the central 10~Mm radius with the averaged unsigned value of the line-of-sight magnetic field within the central 25~Mm radius. We selected this larger radius to capture the active region magnetic field growth up to about 0.5~days after emergence. 
Figure~\ref{fig:avedivflowsm3} (top panel) shows the averaged line-of-sight magnetic field at the surface as it increases during the active region emergence. Although we approximately corrected for the line-of-sight dependence assuming that the surface magnetic field is radial (see Sect.~\ref{sect:avemaps}),  the limitations of this correction can be seen in the increasing flux of the control regions (dashed lines) as they rotate from the limb towards disk centre.
All active regions begin as small bipole regions, and as the flux continues to emerge the bipoles become larger. Because we  limited the area within which we averaged the flux, the active region becomes larger than the averaging area and the flux plateaus at about 1~day after the emergence time. The inset highlights the bin of active regions with pre-emergence bipoles (light green) which evolve to be lower flux active regions, consistent with the results in \cite{AlleySchunker2023}, and is discussed further in Sect.~\ref{sect:bevol}. 

Figure~\ref{fig:avedivflowsm3} (middle panel) shows the evolution of the mean flow divergence centred on the emergence location for the four bins of emerging active regions, as we  defined  in Sect.~\ref{sect:classflows}. The four bins show a consistent flow divergence offset from one another up to two days pre-emergence. The corresponding flow divergence maps for each time interval are shown in the figures in Appendix~\ref{app:fullmaps}. The bin with the strongest converging flows (light green), shows the flow divergence becoming more negative (more convergence) up to \timthreetau and then increasing until the emergence time. Whereas, the bin of EARs with the strongest divergence (dark green) continually increases until the emergence time. The mean diverging flows are all the same at $-2$~days and at 1~day, again reflecting supergranulation lifetimes.

The dashed curves in the middle panel of Fig.~\ref{fig:avedivflowsm3} shows the equivalent evolution of the mean flow divergence  at the centre of the control region maps (Figs.~\ref{fig:avedivmapsqo1}-~\ref{fig:avedivmapsqo4}). The strongest flow divergence values presumably occur close to the centre of supergranules (dark green dashed curve), and the weakest flow divergence (light green dashed curve) presumably occurs in converging lanes between supergranules. The control regions (dashed curves) are consistently positively offset from their emerging active region  counterpart (solid curves).

We subtracted the control region flow divergence from the EARs flow divergence, $\Delta \, \mathrm{div} \, \mathbf{v}_h = \langle \divh \rangle_{r,\mathrm{EAR}} - \langle \divh \rangle_{r,\mathrm{CR}}$ , giving the evolution of the flows in the bottom panel of Fig.~\ref{fig:avedivflowsm3}. The difference in the  flow divergence shows a remarkably consistent profile for each bin, with a converging flow 0.5-1.5~days before emergence. This corresponds to a flow amplitude of about 20\ms, in agreement with \cite{Birchetal2019} and \cite{Gottschlingetal2021}. There is no significant post-emergence divergence flow apparent.

\subsection{Distribution of average flow divergence}

\begin{figure}
	\includegraphics[width=0.5\textwidth]{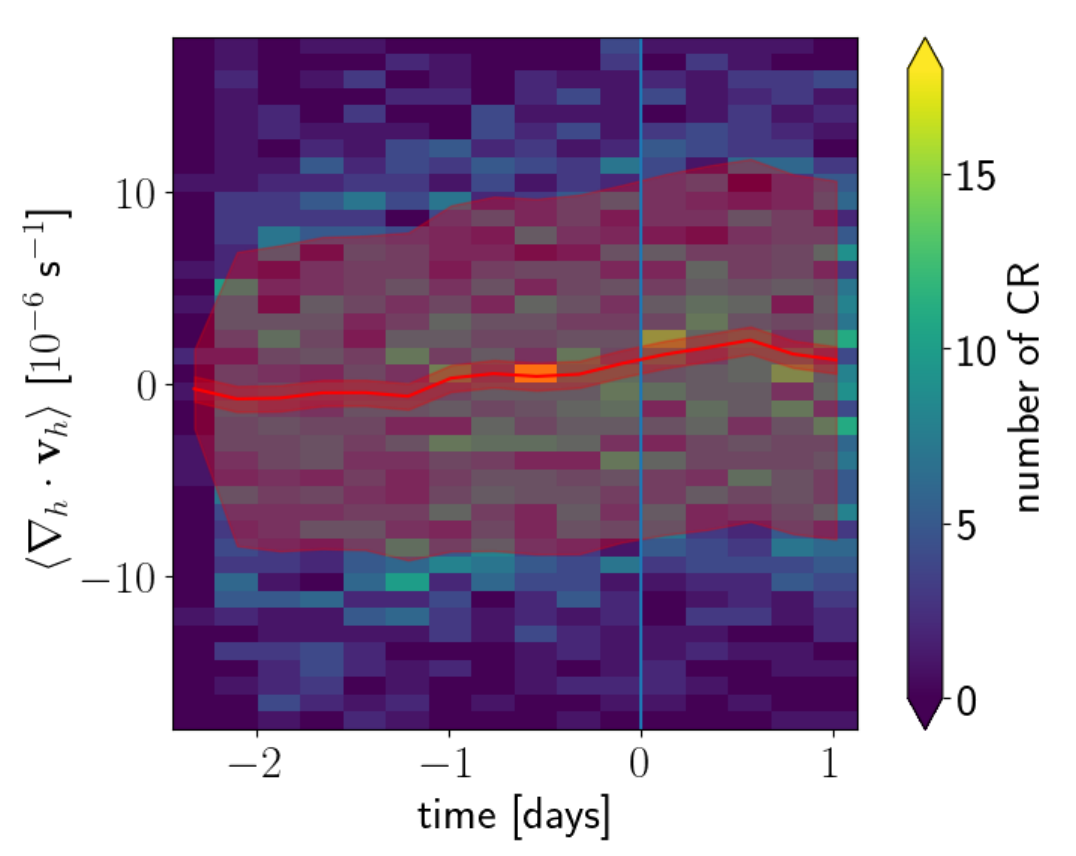}
    \includegraphics[width=0.5\textwidth]{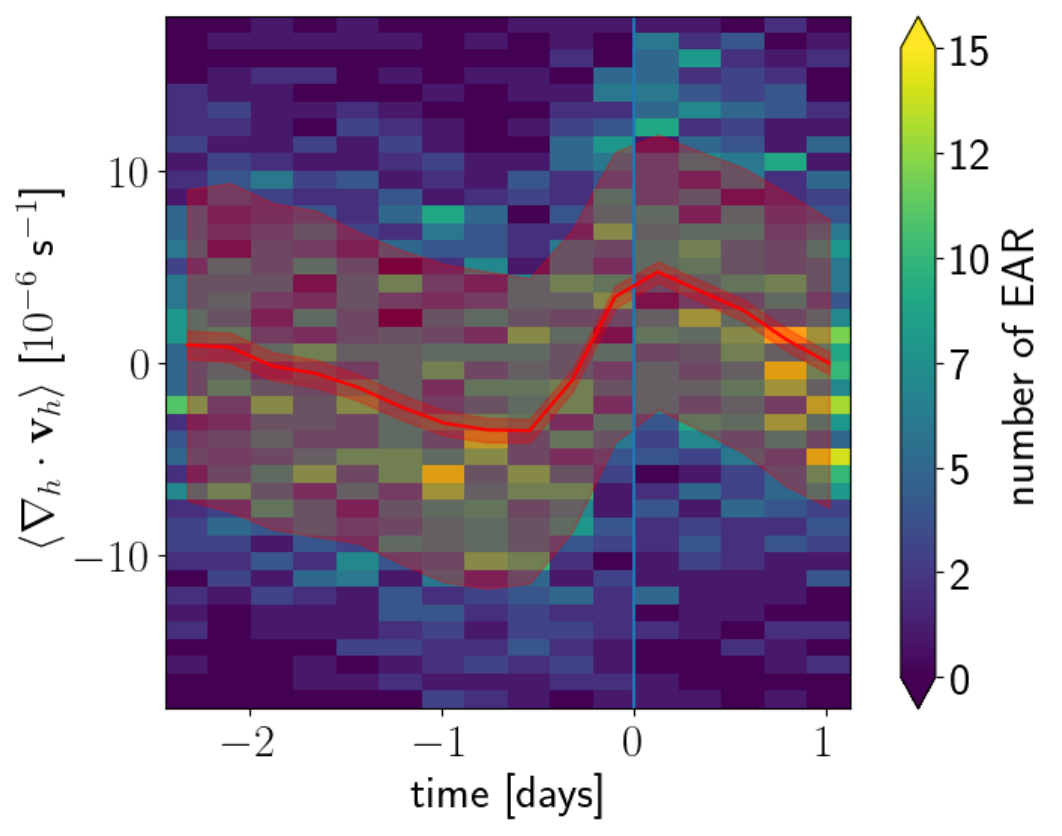} 
    \caption{Distribution of flow divergence averaged over the central 10~Mm radius from the location of emergence in the flow maps as a function of time for the control regions (top) and the EARs (bottom). The red line follows the average and the broad shaded red indicates the standard deviation, and the narrow shaded region is the standard error.  }
    \label{fig:voihistoti}
\end{figure}

We found that the distribution of the flow divergence for all EARs is roughly Gaussian  at all time intervals (Fig.~\ref{fig:voihistoti}),  showing that the converging flow is not due to a single active region.  The observed variation in the flow divergence for the EARS is significant compared to the control regions. The net  flow convergence about 0.5-1~day prior to emergence, and moderate  flow divergence post emergence is shown by the mean (solid red line). This is consistent with the average converging flows reported in \cite{Birchetal2019} prior to emergence.

\section{Dependence on maximum magnetic flux}\label{sect:bevol}

\subsection{Relationship to active regions with pre-emergence bipoles}

\cite{AlleySchunker2023} used the HEARS to classify two samples of emerging active regions based on the presence of magnetic bipoles up to two days prior to emergence. They identified 42 active regions with clear pre-emergence bipoles, and forty-two active regions that emerged abruptly at the emergence time.
 They found that, on average, active regions with persistent pre-emergence bipoles develop into lower magnetic flux active regions post-emergence, and active regions that emerge abruptly develop into higher magnetic flux regions. They also found that active regions that emerge abruptly are associated with strong diverging flows at emergence, and that the EARs with persistent bipoles before emergence are associated with converging flows prior to emergence. 

\begin{figure}
	\includegraphics[width=0.5\textwidth]{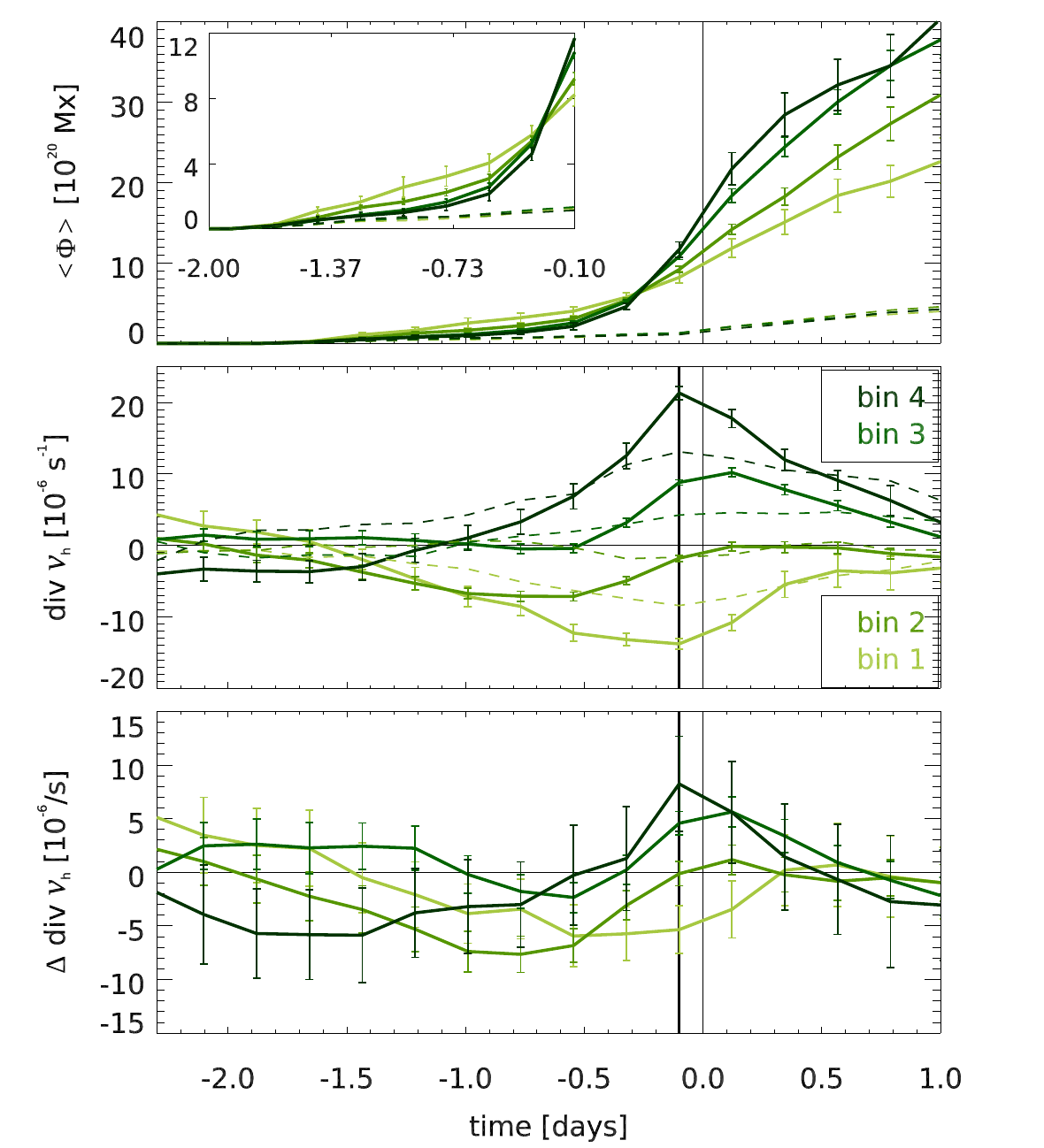}
    \caption{Mean magnetic flux, $\Phi$, and  flow divergence during the emergence process, where $\mathrm{div} \mathbf{v}_h = \langle \divh \rangle_{r,\mathrm{EAR}}$. The EARS were divided into four bins based on the value of the mean divergence in the central 10~Mm radius at \tizerotau (\texttt{TI+00}) indicated by the colour gradient of the curves, from the strongest convergence regions in light green (bin 1) through to the strongest divergence regions in dark green (bin 4).  The thick vertical black line indicates the time at which the bins were defined. The top panel shows the mean flux (note: averaged over the central 25~Mm radius).  The inset shows that the regions with the strongest pre-emergence converging flows have the strongest  pre-emergence magnetic flux. The middle panel shows the mean divergence of the surface flows.  The solid lines are for the emerging active regions and the dashed lines are for the corresponding control regions. The bottom panel shows the \rev{difference, $\Delta \, \mathrm{div} \, \mathbf{v}_h = \langle \divh \rangle_{r,\mathrm{EAR}} - \langle \divh \rangle_{r,\mathrm{CR}}$},  in average divergence between the control regions and active regions as shown in in the middle panel. \rev{The uncertainties are the standard error over all active regions in each bin at each time interval.}
    The converging flow prior to emergence is still evident, however it is not as consistent between the bins as for Fig.~\ref{fig:avedivflowsm3}.}
    \label{fig:avedivflows}
\end{figure}

 Based on the finding that these two samples of EARs identified in \cite{AlleySchunker2023} have distinct flow divergence signals at the time of emergence, we repeated the analysis presented in the first part of this paper (up to Fig.~\ref{fig:avedivflowsm3}) by classifying the flows at \tizerotau (\texttt{TI+00}), when the emergence is underway (instead of at \timthreetau).  
 Figure~\ref{fig:avedivflows} shows that the magnitude of the converging flows in the day prior to emergence is still evident, but the timing is not consistent between the four bins. The higher flux active regions show a converging flow around 1.5 to 2~days prior to emergence, and the lower flux active regions less than 0.5~days before emergence.
The four bins also show a more distinct difference in the magnetic flux evolution (see top panel of Fig.~\ref{fig:avedivflows}) and suggests that the two samples of EARs presented in \cite{AlleySchunker2023} could constitute the extremes of a continuum. 

To address the possibility that the sample identified by \cite{AlleySchunker2023} were the extremes of a continuum, we  excluded these forty-two active regions from our sample and repeated our analysis in the first part of the paper (up to Fig.~\ref{fig:avedivflowsm3}). We found that this did not significantly change the value of the pre-emergence converging flow signal in Fig.~\ref{fig:avedivflowsm3}, suggesting that the signal we observe in this paper from the entire HEARS is not due only to these active regions with pre-emergence bipoles, and they are not the extremes of a continuum.

\subsection{Classification of active regions based on maximum magnetic flux}

\begin{figure}
	\includegraphics[width=0.5\textwidth]{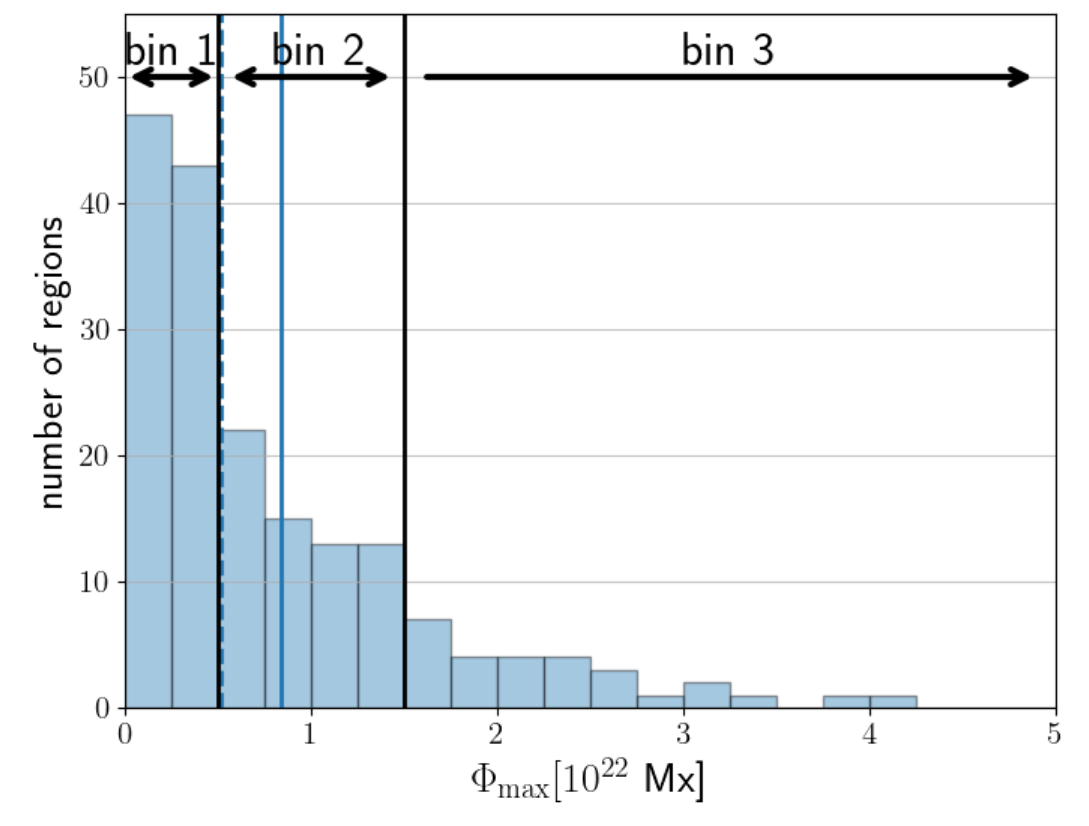}
 \vspace{-0.5cm}
    \caption{Distribution of maximum \texttt{USFLUX} for all EARs. The solid blue line is the mean of the sample and the dashed blue line is the median.  The solid black lines indicate the separation of the three bins (at 0.5$\times 10^{22}$~Mx and 1.5$\times 10^{22}$~Mx).}
    \label{fig:bhisto}
\end{figure}

Motivated by this statistical dependence of the flow divergence on maximum magnetic flux, to test the inverse relationship, we binned the EARs by the maximum unsigned flux, and then computed the averaged flow divergence as a function of time.  We retrieved the maximum unsigned flux of each active region from the Space-weather HMI Active Region Patches, \citep[SHARPS,][]{Bobraetal2014} data series (\texttt{hmi.sharps\_720s}), specifically the \texttt{USFLUX} keyword containing the value of the total unsigned flux computed from the radial component of the vector magnetic field. This is the maximum flux that the active region reaches on the visible solar disk. Figure~\ref{fig:bhisto} shows the distribution of maximum flux and the division into three maximum flux bins. The classification values of $0.5 \times 10^{22}$~Mx and $1.5 \times 10^{22}$~Mx were selected by inspection to span the range and to have enough regions in the high flux bin ($> 1.5 \times 10^{22}$~Mm) to result in a reasonable background noise.

\begin{figure*}
	\includegraphics[width=0.88\textwidth]{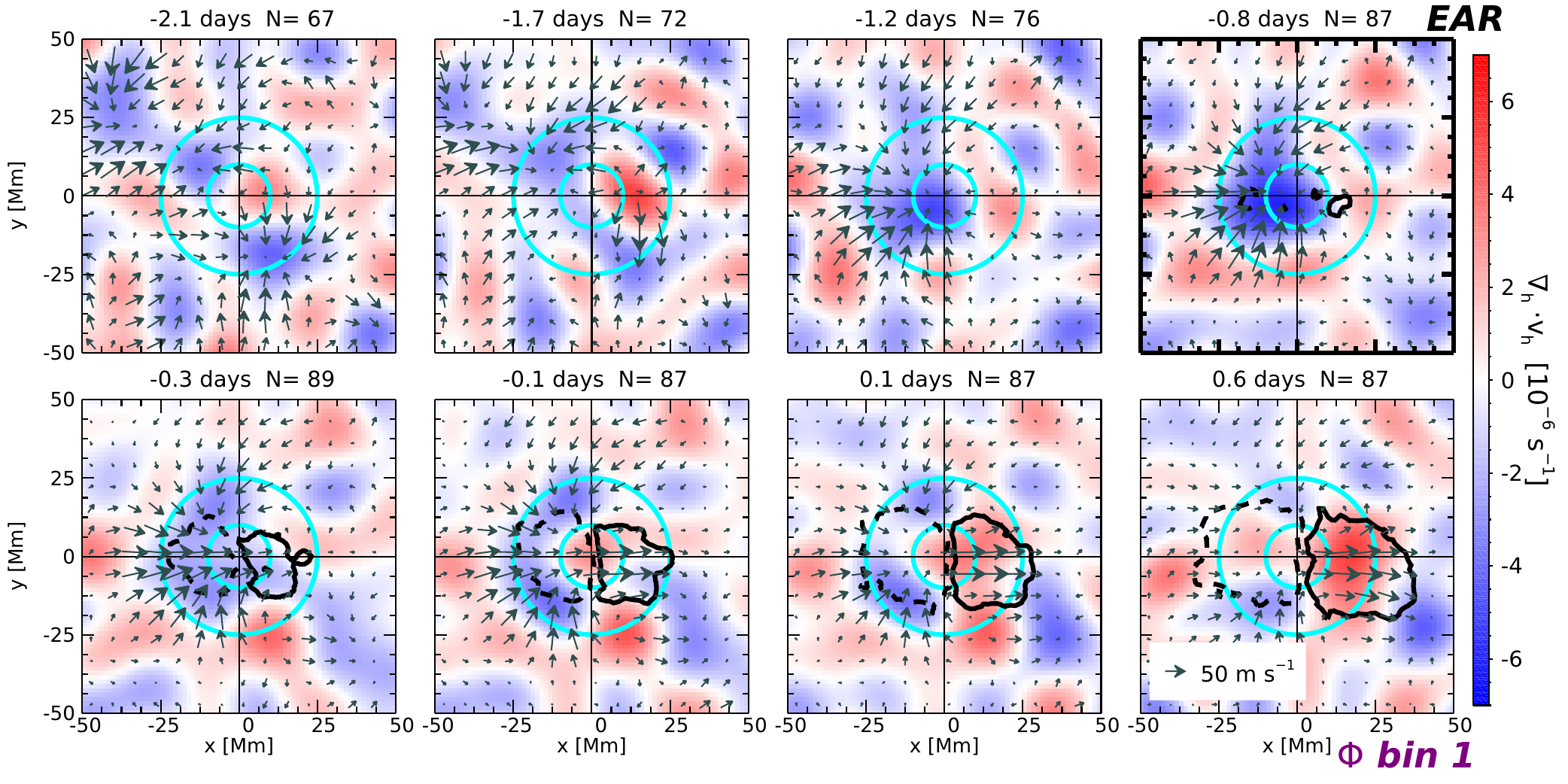}
    \caption{Averaged divergence flow maps for emerging active regions with  $\Phi_\mathrm{max} < 0.5 \times 10^{22}$~Mx (EAR, B bin 1). 
    The arrows indicate the magnitude and direction of the horizontal flows. The solid black contour represents $+20$~G and the dashed contour $-20$~G. The inner cyan circle outlines the area within which the flow divergence is averaged and the outer circle outlines the area within which the absolute magnetic field is averaged. The number of maps contributing to the average is the number $N$. The maps correspond to every second time interval \texttt{TI-09, TI-07, TI-05, TI-03, TI-01, TI+00, TI+01, TI+03.}
    }
    \label{fig:avedivmapb1}
\end{figure*}

\begin{figure*}
	\includegraphics[width=0.9\textwidth]{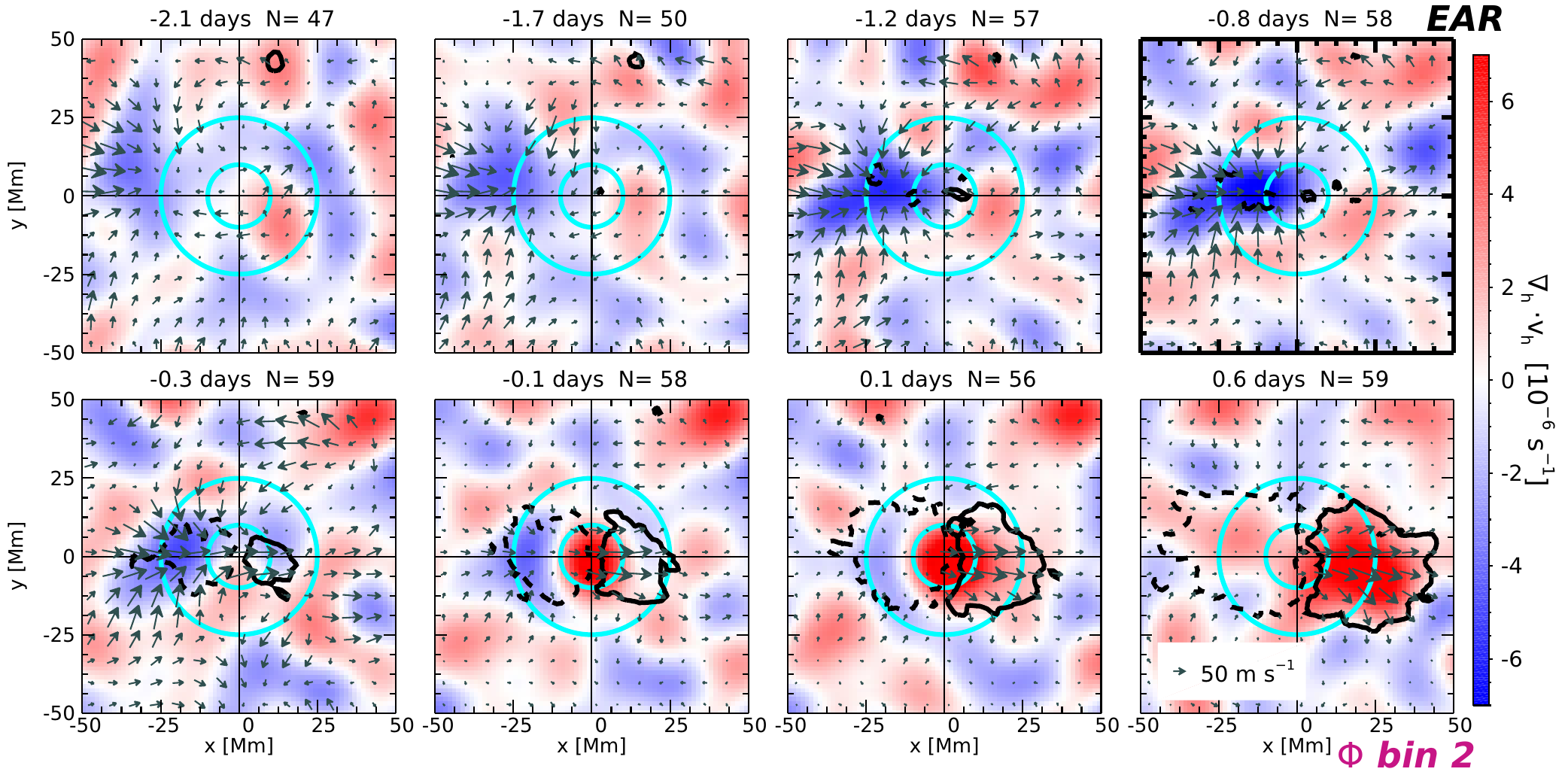}
    \caption{Averaged divergence flow maps for emerging active regions with  $0.5 \times 10^{22} \le \Phi_\mathrm{max} < 1.5 \times 10^{22}$~Mx (EAR, B bin 2).
     The arrows indicate the magnitude and direction of the horizontal flows. The solid black contour represents $+20$~G and the dashed contour $-20$~G. The inner cyan circle outlines the area within which the flow divergence is averaged and the outer circle outlines the area within which the absolute magnetic field is averaged. The number of maps contributing to the average is the number $N$. The maps correspond to every second time interval \texttt{TI-09, TI-07, TI-05, TI-03, TI-01, TI+00, TI+01, TI+03.}
     }
    \label{fig:avedivmapb2}
\end{figure*}

\begin{figure*}
	\includegraphics[width=0.9\textwidth]{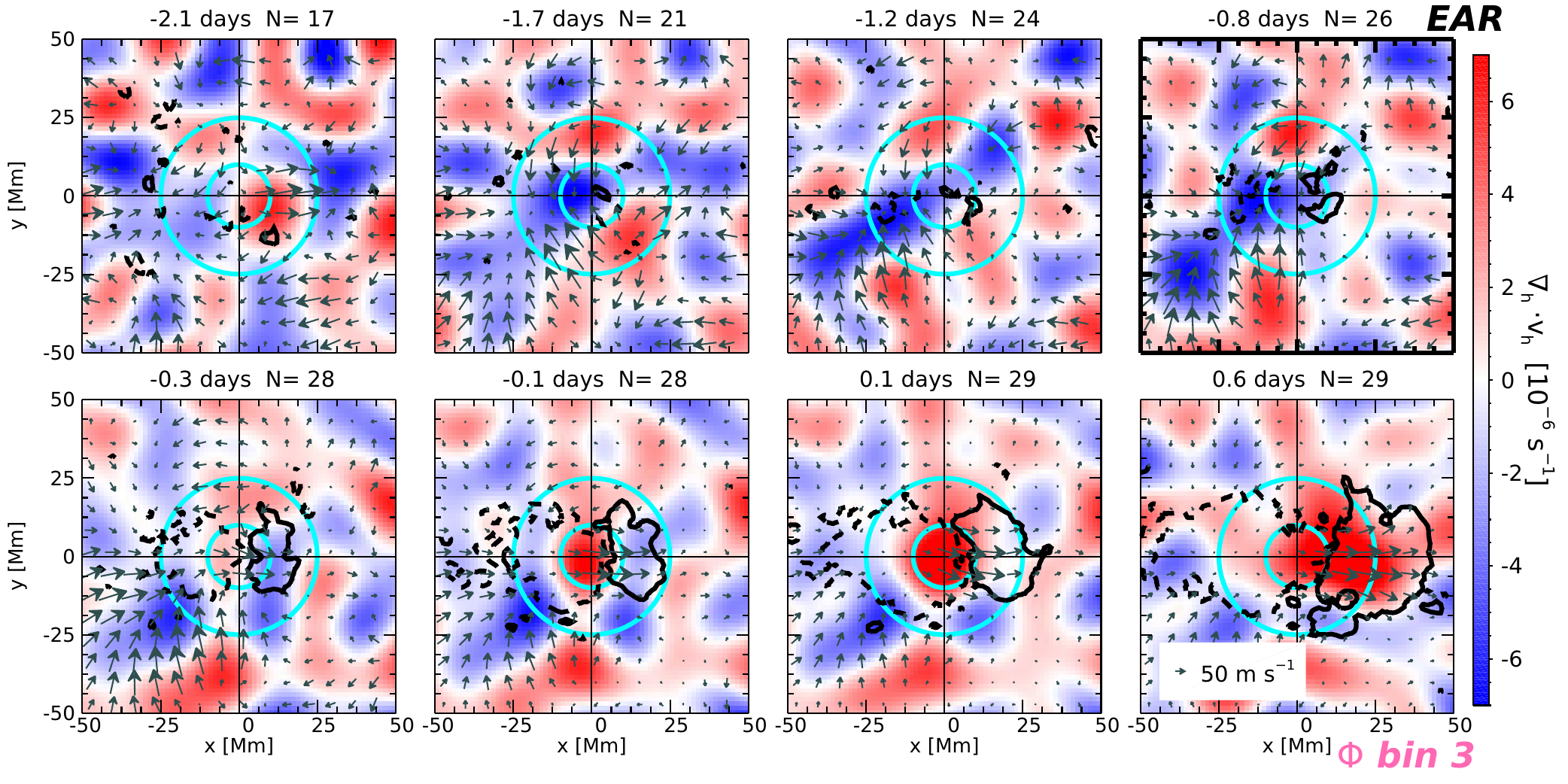}
    \caption{Averaged divergence flow maps for emerging active regions with  $\Phi_\mathrm{max} \ge 1.5 \times 10^{22}$~Mx (EAR, B bin 3).
     The arrows indicate the magnitude and direction of the horizontal flows. The solid black contour represents $+20$~G and the dashed contour $-20$~G. The inner cyan circle outlines the area within which the flow divergence is averaged and the outer circle outlines the area within which the absolute magnetic field is averaged. The number of maps contributing to the average is the number $N$. The maps correspond to every second time interval \texttt{TI-09, TI-07, TI-05, TI-03, TI-01, TI+00, TI+01, TI+03.}
     }
    \label{fig:avedivmapb3}
\end{figure*}

\begin{figure}
\hspace{-0.5cm}
	\includegraphics[width=0.5\textwidth]{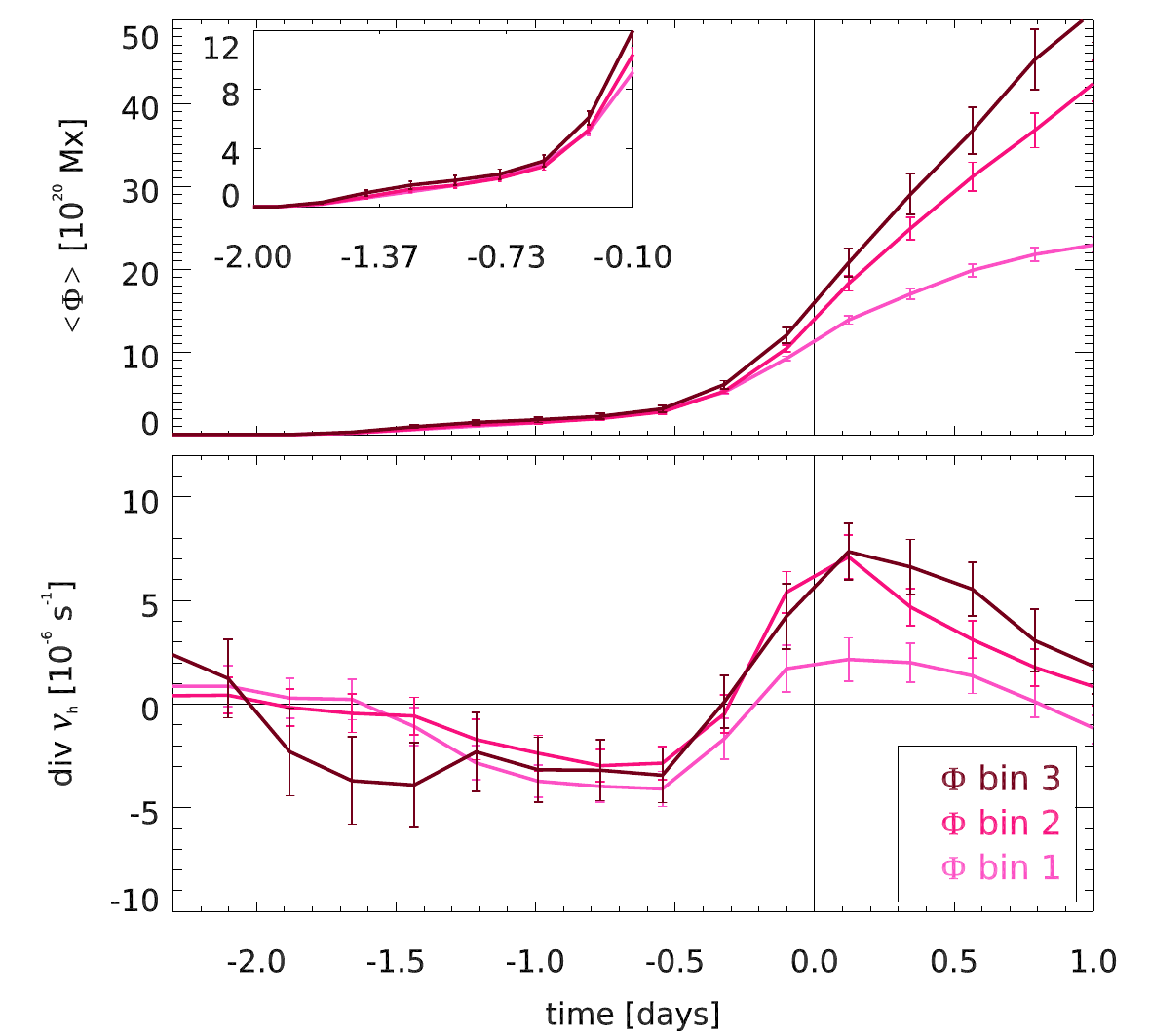}
 \vspace{0cm}
    \caption{Mean line-of-sight unsigned magnetic flux ($\langle \Phi \rangle_{r,\mathrm{EAR}}$, in the central 25~Mm radius, top panel) and mean divergence flows ($\mathrm{div} \, v_h = \langle \divh \rangle_{r,\mathrm{EAR}}$, bottom panel) during the emergence process for EARS divided into three bins based on the value of the  maximum flux from lowest flux bin ($\Phi_\mathrm{max}<0.5\times 10^{22}$~Mx, light pink) to highest flux bin ($\Phi_\mathrm{max} \ge 1.5\times 10^{22}$~Mx, dark pink).    
    }
    \label{fig:avedivflowsb}
\end{figure}

\begin{figure}
    \includegraphics[width=0.45\textwidth]{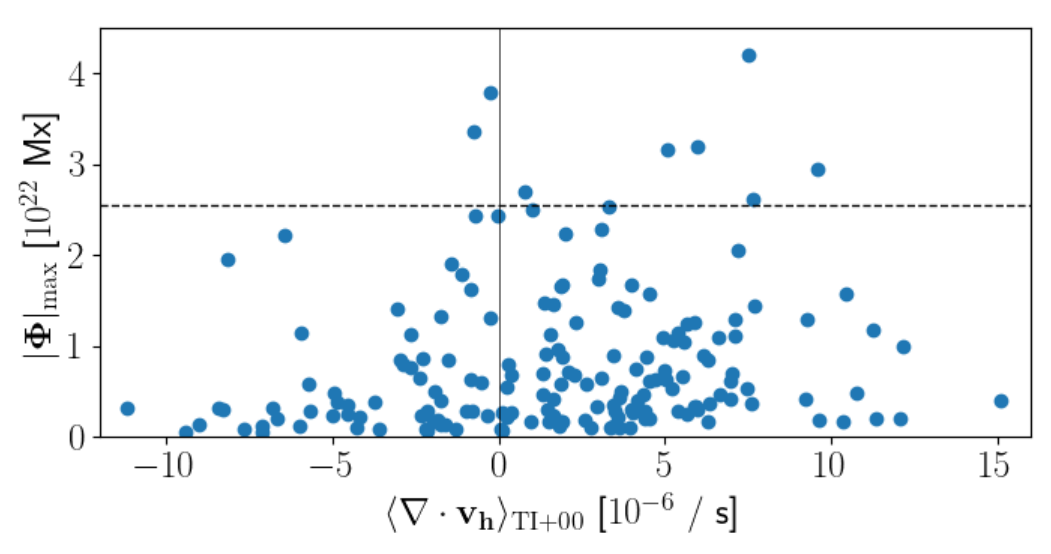}
    \caption{Maximum unsigned flux of the active regions as a function of the average flow divergence at \texttt{TI+00} (\tizerotau). Six of the eight regions with highest maximum flux (above the dashed horizontal line to guide the eye) in the sample considered here have a diverging flow at the emergence time and location. }
    \label{fig:divmaxb}
\end{figure}

Figures~\ref{fig:avedivmapb1}-\ref{fig:avedivmapb3} show that when EARs are binned by $\Phi_\mathrm{max}$, they have  consistent converging flows prior to emergence, followed by a flux dependent diverging flow at the time of emergence concentrated mostly in the leading polarity.
Figure~\ref{fig:avedivflowsb} shows the evolution of the flow divergence maps averaged over the central 10~Mm radius, with a mean horizontal flow divergence $\langle \nabla \cdot \mathbf{v}_h\rangle_{r,\mathrm{EAR}} = -3.7 \pm       2.2 \times 10^{-6}$s$^{-1}$ at $-0.7$~days.
Assessing the individual active regions (see Figure~\ref{fig:divmaxb}) confirms a weak tendency for higher magnetic flux active regions to emerge with stronger diverging flows.

\section{Simulation of bipole formation in convective flows}\label{sect:sims}

\rev{Statistically, we have shown that active region emergence is correlated with an increase of the horizontal flow divergence at the surface, starting about a day before emergence. This increase is superposed on flows with a longer lifetime (see Fig.~\ref{fig:avedivflowsm3}), and with velocities comparable to those of the convection. To qualitatively interpret the observations we aim to simulate the interaction of a flux tube with the convective flows as it rises through the near-surface layers of the Sun.}

\rev{There are a number of ways to initiate flux emergence through simulations of the near-surface convection zone, e.g. forcing a semitorus of flux through the bottom boundary \citep{Cheungetal2010}, with \citep{RempelCheung2014} and without \citep{Fanetal2003} a twisted magnetic field, and a passive flux tube placed in the subsurface convection \citep{HottaIijima2020}.
From these, the emergence in \citet{Cheungetal2010} and \citet{RempelCheung2014} produce an outflow as it emerges at the surface, whereas the emergence in \citet{HottaIijima2020} does not produce a significant outflow. 
\cite{Birchetal2016} demonstrated that the outflow is directly associated with the rise speed of the flux tube. None of the observed active regions in the HEARS sample produced a significant outflow, and \citet{Birchetal2016} placed an upper limit on the rise speed of a flux tube from a depth of $20$~Mm of $150$~\ms. For this reason, we aimed to simulate an emerging active region that did not produce significant surface outflows.}

\rev{We began with a pre-computed fully hydrodynamical simulation, computed using the MURaM code \citep{Voegleretal2005, Rempel2017}, in a box of size $36.864 \times 18.432 \times 16.128 $~Mm  ($384 \times 192\times 504$~grid points), extending up to about $15$~Mm below the surface. To study flux emergence on this timescale, we imposed a magnetic flux tube at a depth of $11$~Mm. Placing the tube at about 4~Mm from the bottom of the box limits the width of the imposed flux tube. We chose a FWHM of 1.3~Mm, and, to ensure a reasonable flux ($10^{21}$~Mx), we chose a peak field strength of $50$~kG (Appendix~\ref{app:icsims} fully describes the simulation domain).}

\begin{figure}
	\includegraphics[width=0.5\textwidth]{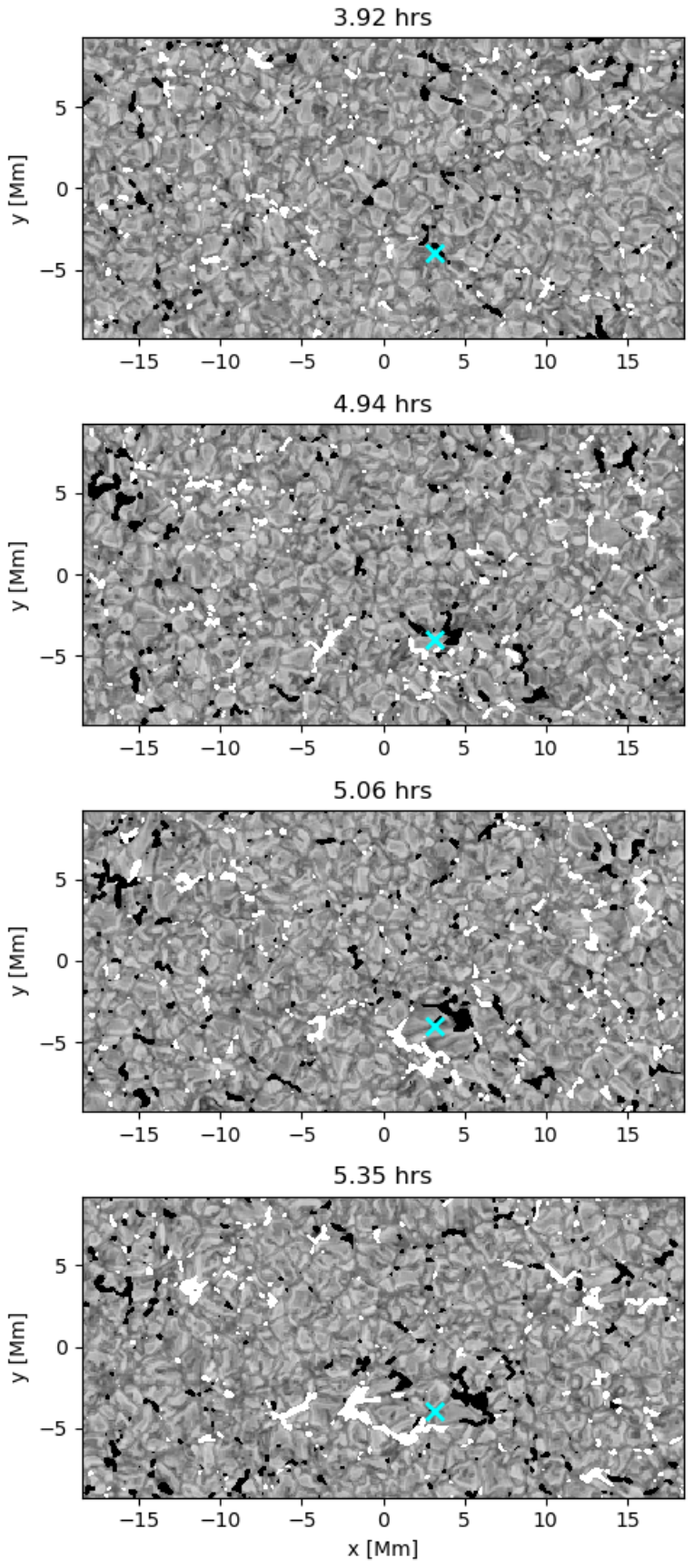}
    \vspace{-0.7cm}
    \caption{Surface intensity at different times since the beginning of the MURaM simulation showing the granulation, normalised to the mean of the intensity of the top panel (greyscale ranges from 0 to 2). The vertical magnetic field at the surface (black and white filled contours for $\pm 1000$~G) shows a well-defined bipole feature  at 5.35~hours. The cyan cross indicates the emergence location of the bipole ($(x,y)=(3.12, -3.93)$~Mm) which was defined by inspection of the vertical magnetic field maps.}
    \label{fig:bpsurf}
\end{figure}

\rev{The flux tube initially rose towards the surface due to the lower density. Above about $-8$~Mm in height the downflows in the convection were sufficient to counter the rise so that the tube formed arch-like structures (see Fig.~\ref{fig:ictube}, Appendix~\ref{app:icsims}), in reasonably strong convective upflows consistent with the results of \citep[e.g.][]{SteinNordlund2012}. Ultimately  a bipolar feature was formed at the surface after about 5~hours (see Fig.~\ref{fig:bpsurf}). }

\begin{landscape}
  \begin{figure}
	\includegraphics[scale=0.65]{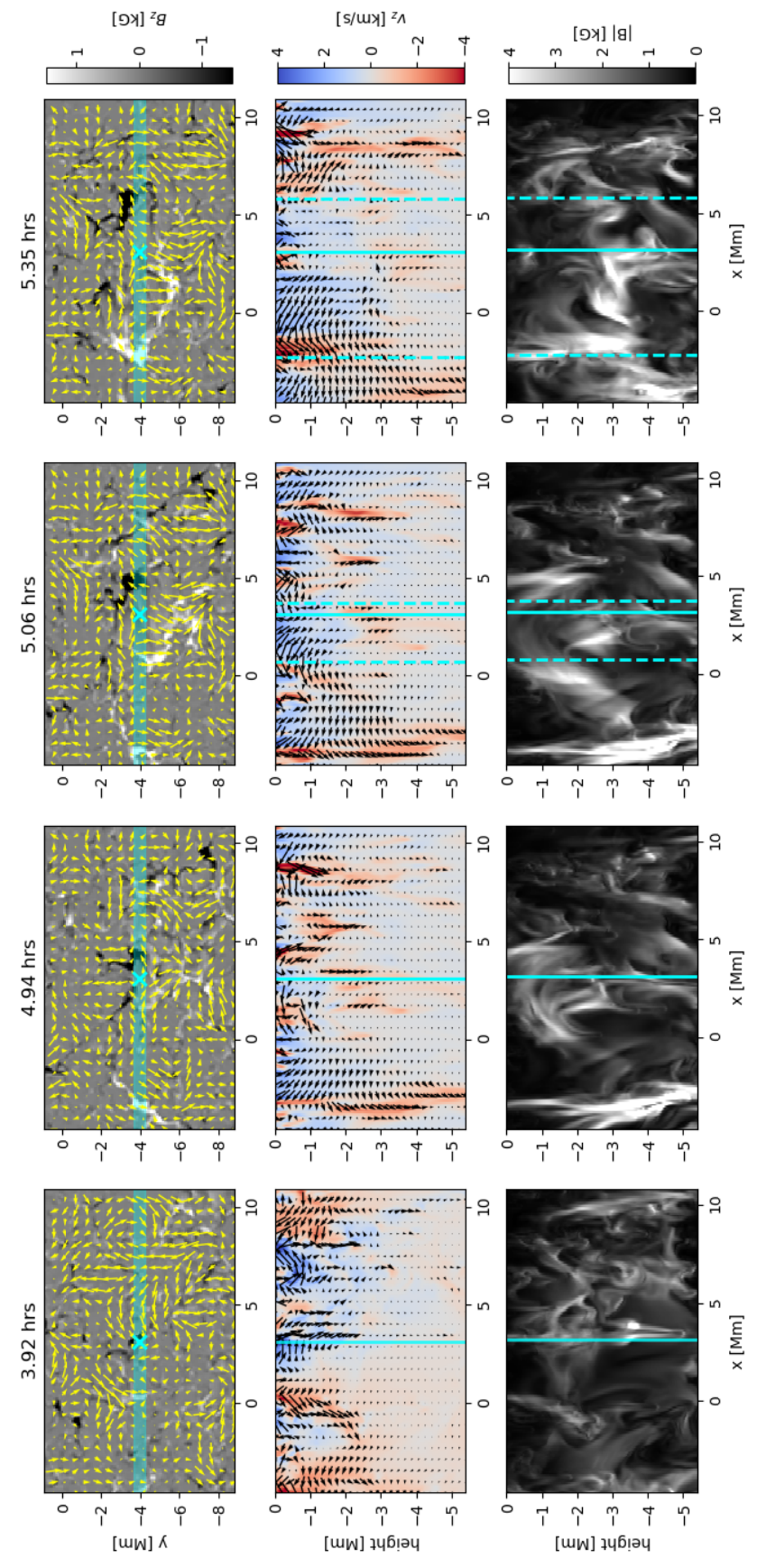}\\
       \caption{Evolution of the formation of the surface bipole. Time, since the beginning of the simulation, increases across columns from left to right. The top row shows the vertical magnetic field at the surface cropped and centred on the bipole. The yellow arrows indicate the magnitude and direction of the horizontal flows at the surface. The cyan cross indicates the emergence location of the bipole ($(x,y)=(3.12, -3.93)$~Mm, identified by eye). The second row shows the vertical velocity (red downflows, and blue upflows) below the surface averaged over the cyan strip in the top row. The black arrows indicate the direction of the horizontal flows in the vertical plane. The bottom row shows the absolute magnetic field below the surface, where a rising arch of flux tube can be seen forming the bipole. The solid cyan line indicates the sub-surface location of the emergence, and the dashed lines represent the sub-surface location of the maximum and minimum values of the surface magnetic field in the last two frames. }
            \label{fig:bpflows}     
    \end{figure}
\end{landscape}

 The bipole formed with diverging flows on the order of the convective velocities at the surface (top row, Fig.~\ref{fig:bpflows}).  Below the surface, an arc of flux rose towards the surface (bottom row, second column, Fig.~\ref{fig:bpflows}) in an upflow region  2-3~Mm below the surface, and the legs were in downflow regions either side (middle row, second, third and fourth column Fig.~\ref{fig:bpflows}).
The flux was then swept outwards with the diverging flows at the surface and formed concentrated polarities in converging downflows (top row, Fig.~\ref{fig:bpflows}). 
\rev{This picture is consistent with previous results of more passive flux emergence simulations such as \citep{SteinNordlund2012,HottaIijima2020}.} We also note that the separation of the polarities is on the order of the spatial scale of the convective flows bringing it to the surface, consistent with observations \cite{Schunkeretal2020}. 

\rev{Despite the enhanced buoyancy of the $50$~kG tube, only about $10$~\% of the magnetic flux in the tube (initially $10^{21}$~Mx) actually emerges. This suggests that even having a tube with a strong initial magnetic field (albeit with low flux), and in a large-scale upflow is not sufficient to ensure a substantial fraction of the flux emerges from a depth of 11~Mm.  }

\rev{Our simulation of an emerging bipole is encouraging, however, this bipole does not develop into a stable active region as observed on the Sun. We emphasise that the converging flow prior to emergence, presented in the previous sections, is only observable after averaging over the intrinsic convective flows. 
To identify these low-amplitude flows in the simulations, it would be necessary to average over a similar number of emergences.}

\newpage
\section{Discussion}\label{sect:disc}

\begin{figure*}
\includegraphics[width=0.9\textwidth]{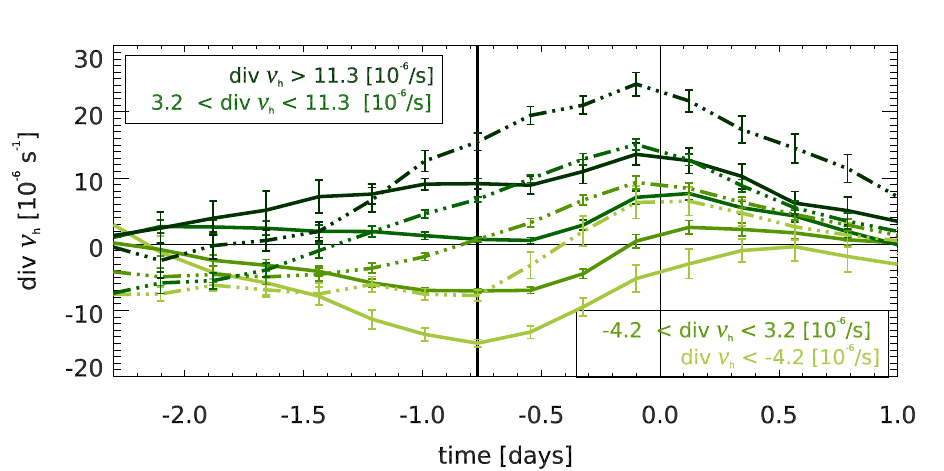}
    \caption{Mean flow divergence over emerging active regions in each  bin (solid curves, the same as in Fig.~\ref{fig:avedivflowsm3}), and mean flow divergence over forming supergranules in the control regions in each bin (dash-dot curves). The divergence profiles of emerging active regions and forming supergranules are similar. The legend indicates the bin  values for the control regions. From the control sample, we selected regions where the maximum spatially averaged flow divergence $\langle \nabla_h \cdot v_h \rangle_r$ between TI-20 and TI+20, was also a local maximum turning point and was continually increasing in the three time intervals ($\approx$16~hours) prior to the time of the identified maximum. This resulted in 91 control regions included in this sample. We defined shifted time intervals re-assigning TI+00 to the time of the identified local maximum, in an attempt to model the flow divergence profiles observed for the EARs in Fig.~\ref{fig:avedivflowsm3}. We then treated these flow divergence profiles, as we did in Section~\ref{sect:classflows}: we computed the distribution of flows at the shifted TI-03  and defined four  bins (at $\langle \nabla_h \cdot v_h \rangle_{r,\mathrm{SG}} = -4.2 \times 10^{-6}, \, 3.2 \times 10^{-6}, \, 11.3 \times 10^{-6}$~s$^{-1}$ ); we then averaged the flow divergence profiles,  $\langle \nabla_h \cdot v_h \rangle_{r,\mathrm{SG}}$, over the control regions in each bin.}
    \label{fig:avemaxdivtime}
\end{figure*}

The evolution of the statistical flow divergence in each bin of EARs shows that active regions can emerge in either diverging or converging flows in the supergranulation convection pattern, and that the flow divergence consistently increases in the 0.5-1~day prior to emergence.
This increase in flow divergence is reminiscent of a forming supergranule. To test this, we selected forming supergranules in the control regions, and repeated the main analysis which shows that supergranulation flows in the control regions have in general a higher flow divergence (see Fig.~\ref{fig:avemaxdivtime}). This could be a consequence of the method we used to identify forming supergranules (see caption of Fig.~\ref{fig:avemaxdivtime}), or perhaps active region formation does occur as supergranules form, but the supergranulation flows are superposed on some larger-scale, lower-amplitude, longer-lived flow divergence. 
To explore this, we also repeated the analysis by filtering out flows with scales larger than 70~Mm (or $k R_\odot < 10$) in addition to filtering out noise with scales smaller than 3~Mm (or $k R_\odot > 220$), and repeated the main analysis. This did not result in a significant difference to the flow divergence  profiles, however, this deserves a closer inspection, before we can constrain the spatial scale of the underlying flow divergence.

Our results also point to an active component to the emergence process, with lower flux active regions tending to form in converging flows and higher flux active regions tending to emerge in diverging flows.
The clear differences in the flow divergence of emerging active regions that defines the  bins, extends up to two days before, and one day after, the emergence time.

It appears that, statistically, active regions, regardless of their associated flow divergence or maximum flux, are related to a converging flow superposed on the background convective flows about one day before they emerge. This corresponds to a flow speed of about about 20\ms,
an order of magnitude less than supergranulation flows of 300-500~ms$^{-1}$ \citep{SimonLeighton1964}.
The apparent additional converging flow in the day before emergence may be due to the emerging magnetic flux itself, or it may represent a preference for active regions to emerge in a larger scale converging flow featured in each statistical sample.

Our results support a passive emergence process with convective flows bringing the flux to the surface. 
Although a single case of emergence is not representative of the observed statistical result, we were able to simulate an emerging bipole and place it in the context of the observed sample of active regions.
Our observation that higher flux regions emerge with diverging flows is consistent with the simulated bipole formation. 
In the context of the observations, our simulations suggest that low magnetic flux active regions which grow slowly, are brought up in upflows, and swept outwards into the converging flow lanes. However, they can only be observed once the flux has formed coherent features large enough to be resolved by the instrument. The consequence of this is that we observe small, weak active regions only after they have been swept into the converging flows. Whereas, the larger, high flux regions that emerge more abruptly \citep[as shown in][and supported by our results here]{AlleySchunker2023} can be observed directly at the site of the diverging flows.

In summary, we found that active regions are observed to emerge anywhere in the supergranulation scale convective flow pattern, but that the divergence increases in the day prior to emergence, and lower (higher) flux active regions tend to emerge into converging (diverging) flows. Regardless of magnetic flux, we found that all active regions  are associated with an additional weak converging flow at about $-0.8$~days prior to emergence, which becomes weaker until emergence. 
A statistical sample of simulated emerging active regions of different flux and at different locations in the convective flow pattern would be the ideal experiment to explain the physical relationship between the magnetic flux and convection.

\clearpage
\section*{Acknowledgements}
HS is the recipient of an Australian Research Council Future Fellowship Award (project number FT220100330) and this research is partially funded by this grant from the Australian Government.
HS, WRB and LG acknowledge support from the project ``Preparations for PLATO asteroseismology'' DAAD project 57600926.
DCB is supported by NASA grants 80NSSC20K0187 (Living With a Star program), 80NSSC22K0754 (Heliophysics Guest Investigator program), and 80NSSC22M0162 (COFFIES DRIVE Science Center).
The data were processed at the German Data Center for SDO, funded by the German Aerospace Center under grant DLR 50OL1701.  Observations courtesy of NASA/SDO and the HMI science teams. HS \& WRB acknowledge the Awabakal people, the traditional custodians of the unceded land on which their research was undertaken.
We thank the anonymous referee for a thorough, thoughtful and carefully considered report.

\section*{Data Availability}

The HEARs data can be reproduced following the description in \cite{Schunkeretal2016}. The results presented in this paper are available through private communication with the authors.



\bibliographystyle{mnras}
\bibliography{ears_mnras_ms}



\clearpage

\appendix

\section{NOAA Active Region numbers used to define bins of flow divergence}
List of EARS used to define the bins of flow divergence at \texttt{TI-03} in order of increasing $\langle \divh \rangle_r$. Refer to Table A.1 in \cite{Schunkeretal2016,Schunkeretal2019} for more details about the time and location of emergence. \\
  11154, 
   11833, 
   11574, 
   11902, 
   11624, 
   12039, 
   11098, 
   11802, 
   12089, 
   11807, 
   11962, 
   11702, 
   11103, 
   11194, 
   11318, 
   11198, 
   11992, 
   11182, 
   11811, 
   11829, 
   11813, 
   11626, 
   11130, 
   11640, 
   11786, 
   11143, 
   11437, 
   11138, 
   11385, 
   12048, 
   11607, 
   11116, 
   11300, 
   11551, 
   11381, 
   11894, 
   12029, 
   11070, 
   11631, 
   11136, 
   11397, 
   11167, 
   11400, 
   11148, 
   11510, 
   11645, 
   11603, 
   11416, 
   11874, 
   11206, 
   11406, 
   11712, 
   11146, 
   11911, 
   11141, 
   11946, 
   11199, 
   11675, 
   11122, 
   11849, 
   11627, 
   11511, 
   11291, 
   11322, 
   11750, 
   11821, 
   11242, 
   11152, 
   11781, 
   11297, 
   11523, 
   11370, 
   11114, 
   11157, 
   11570, 
   11137, 
   11223, 
   12118, 
   11776, 
   11686, 
   11831, 
   11932, 
   11886, 
   11076, 
   11159, 
   11549,
   11680, 
   11789, 
   11449, 
   11080, 
   11267, 
   11824, 
   11988, 
   11752, 
   11456, 
   11334, 
   12099, 
   11075, 
   11842, 
   12064, 
   12078, 
   11697, 
   11079, 
   11878, 
   11565, 
   11924, 
   11132, 
   11072, 
   11497, 
   11915, 
   11327, 
   11696, 
   11670, 
   11500, 
   11726, 
   11174, 
   11554, 
   11156, 
   11088, 
   11945, 
   11784, 
   11304, 
   11703, 
   11951, 
   11978, 
   11969, 
   11764, 
   11222, 
   11472, 
   11597, 
   11414, 
   11547, 
   11273, 
   11142, 
   12041, 
   11214, 
   11561, 
   11718, 
   11105, 
   11404, 
   11294, 
   11331, 
   11396, 
   11707, 
   11431, 
   11086, 
   11288, 
   12011, 
   11239, 
   11780, 
   11074, 
   12098, 
   11855, 
   11922, 
   11209, 
   11310, 
   11241, 
   11560, 
   12062, 
   11706, 
   11910, 
   12105, 
   11843, 
   11066, 
   11290, 
   11158, 
   11605, 
   11867, 
   11531, 
   12119, 
   11081, 
   11211.


\clearpage
\section{Flow divergence maps at each time interval}\label{app:fullmaps}
For completeness, here we show the flow divergence maps for each emerging active region bin at each time interval, corresponding to Figs.~\ref{fig:avedivmapso1m3}, \ref{fig:avedivmapso2m3}, \ref{fig:avedivmapso3m3}, and \ref{fig:avedivmapso4m3}.

\begin{figure*}
	\includegraphics[width=0.8\textwidth]{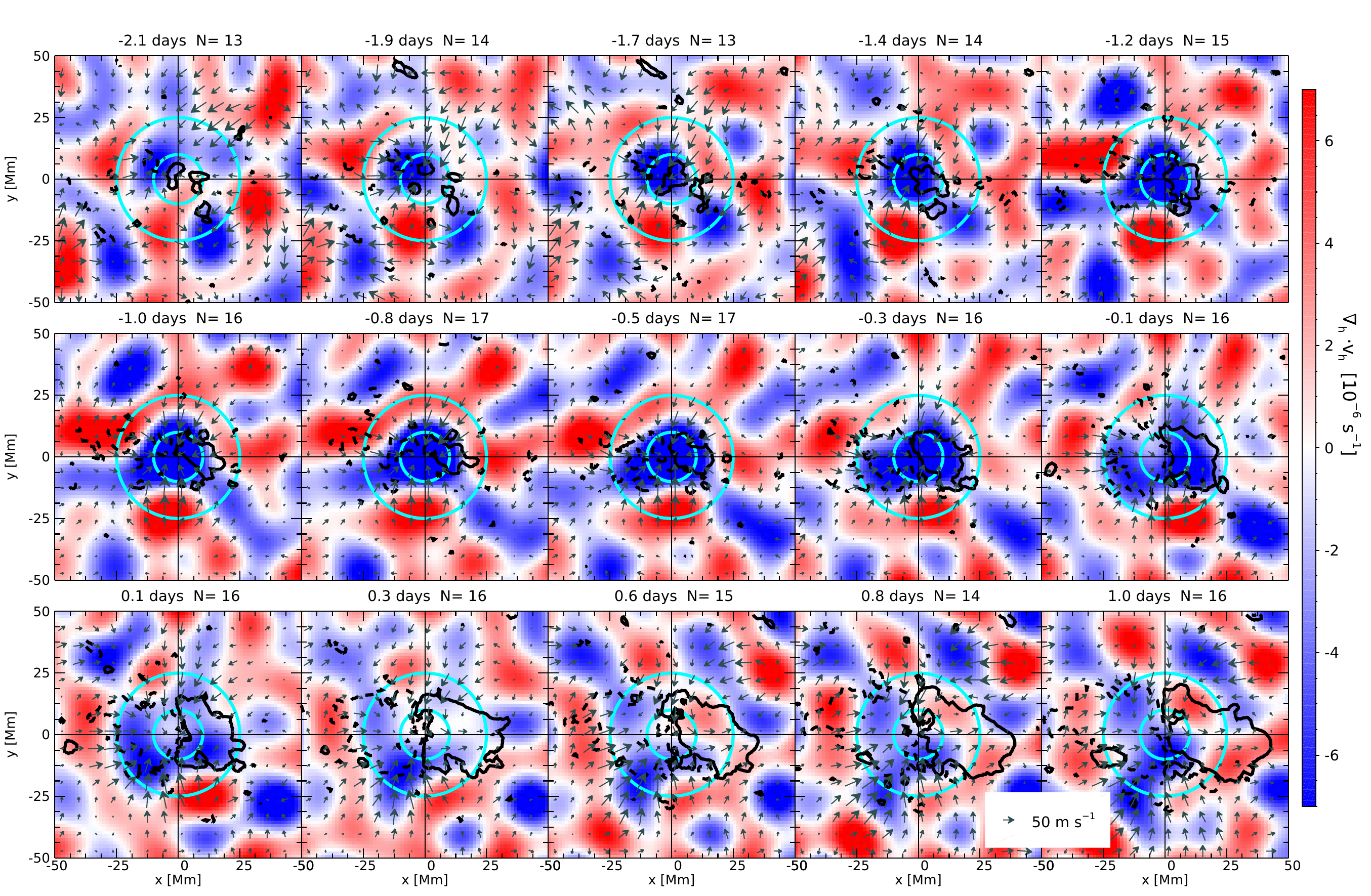}
    \caption{Averaged divergence flow maps for EARs at \texttt{TI-03} or \timthreetau with $\langle \divh \rangle_r \leq -11.4 \times 10^{-6}$~\ps (EAR, bin 1). The arrows indicate the magnitude and direction of the horizontal flows. The solid black contour represents $+20$~G and the dashed contour $-20$~G. The number of maps contributing to the average is the number $N$. The inner cyan circle outlines the area within which the flow divergence is averaged and the outer circle outlines the area within which the absolute magnetic field is averaged. The maps correspond each time interval in the range from \texttt{TI-09} to \texttt{TI+05}.}
    \label{fig:voifullmaps1}
    \end{figure*}

    \begin{figure*}
	\includegraphics[width=0.8\textwidth]{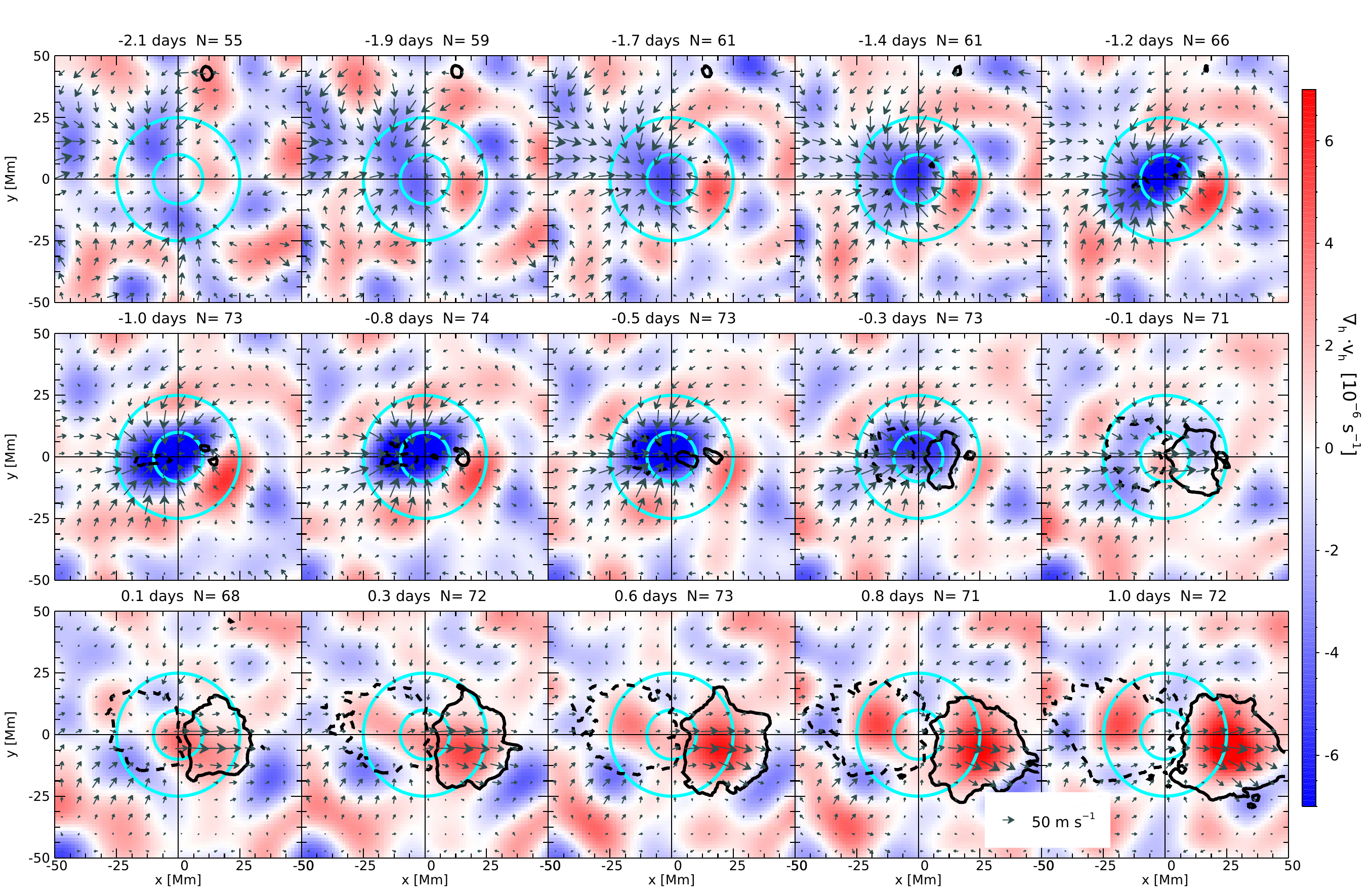}
    \caption{Averaged divergence flow maps for EARs at \texttt{TI-03} or \timthreetau with $-11.4 \times 10^{-6} < \langle \divh \rangle_r \leq -3.5 \times 10^{-6}$~\ps (EAR, bin 2). 
    \rev{All other annotations are the same as Fig.~\ref{fig:voifullmaps1}.}
    }
    \label{fig:voifullmaps2}
    \end{figure*}

\begin{figure*}
	\includegraphics[width=0.8\textwidth]{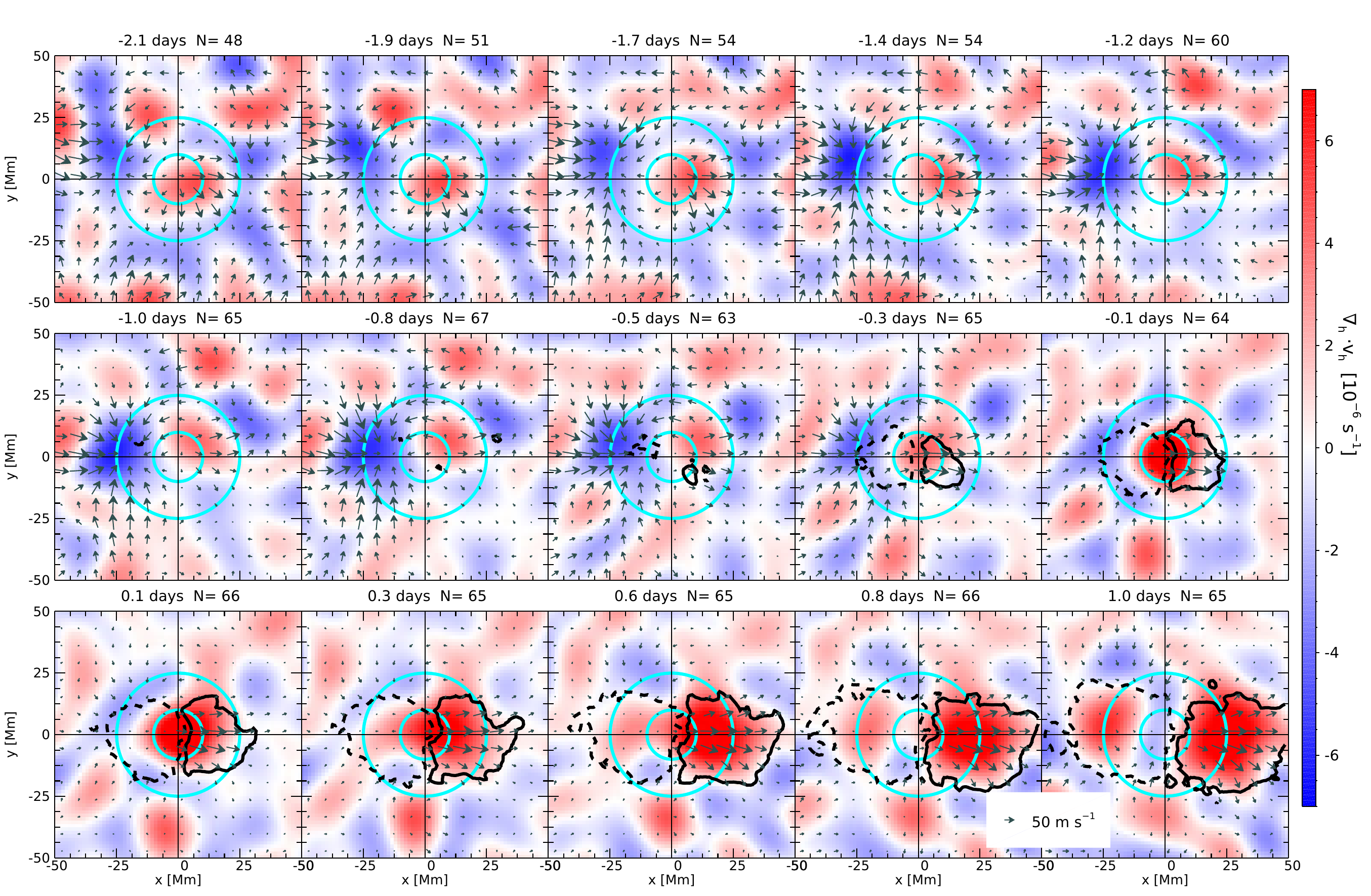}
    \caption{Averaged divergence flow maps for EARs at \texttt{TI-03} or \timthreetau with $ -3.5 \times 10^{-6} < \langle \divh \rangle_r \leq 6.5 \times 10^{-6}$~\ps (EAR, bin 3).
    \rev{All other annotations are the same as Fig.~\ref{fig:voifullmaps1}.}
    }
    \label{fig:voifullmaps3}
    \end{figure*}

    \begin{figure*}
	\includegraphics[width=0.8\textwidth]{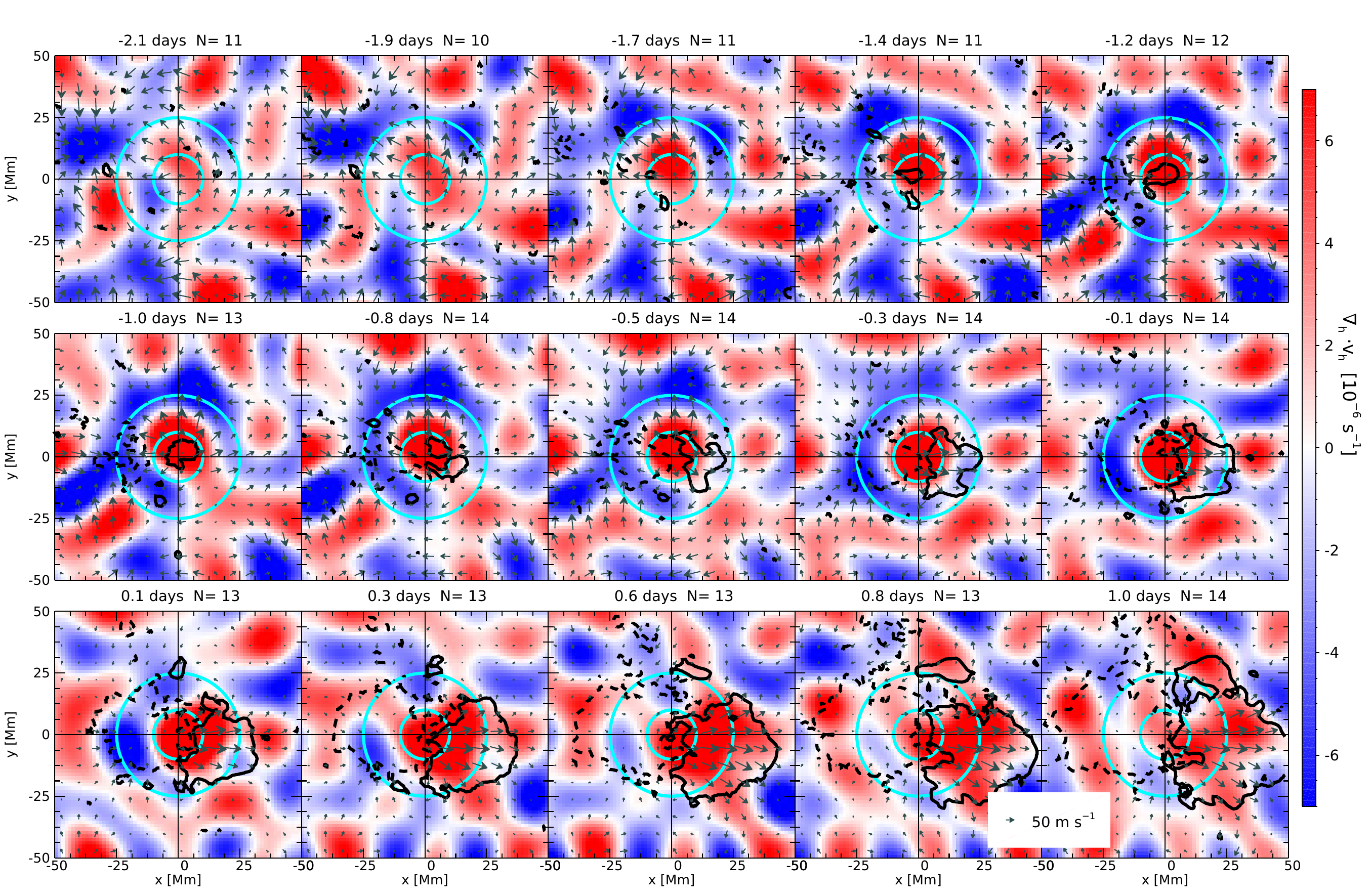}
    \caption{Averaged divergence flow maps for EARs at \texttt{TI-03} or \timthreetau with $\langle \divh \rangle_r > 6.5 \times 10^{-6}$~\ps (EAR, bin 4).     \rev{All other annotations are the same as Fig.~\ref{fig:voifullmaps1}.}
    }
    \label{fig:voifullmaps4}
    \end{figure*}

\clearpage
\section{MURaM simulation domain}\label{app:icsims}

We simulated the emergence of flux with MURaM in a three-dimensional box with a total size of $36.864 \times 18.432 \times 16.128 $~Mm  ($384 \times 192\times 504$~pixels), where the surface was 1~Mm below the top of the box.  We used the upper and lower boundary condition as described in \cite{Voegleretal2005}. The top boundary is a potential field, and the bottom boundary sets the horizontal component of magnetic field at the bottom of the box to zero in order to keep the net vertical magnetic flux in the computational domain height independent and constant in time. We used grey radiative transfer at the surface, which corresponds to a $\tau$-level computed with mean opacities, which does not correspond to a constant geometric height surface. We used the combined equation of state
\citep[from OPAL and Uppsala Opacity Package][]{Rogersetal1996,Gustafsson1975} as described in \cite{Rempel2017}.

\begin{figure*}
	\includegraphics[width=1.\textwidth]{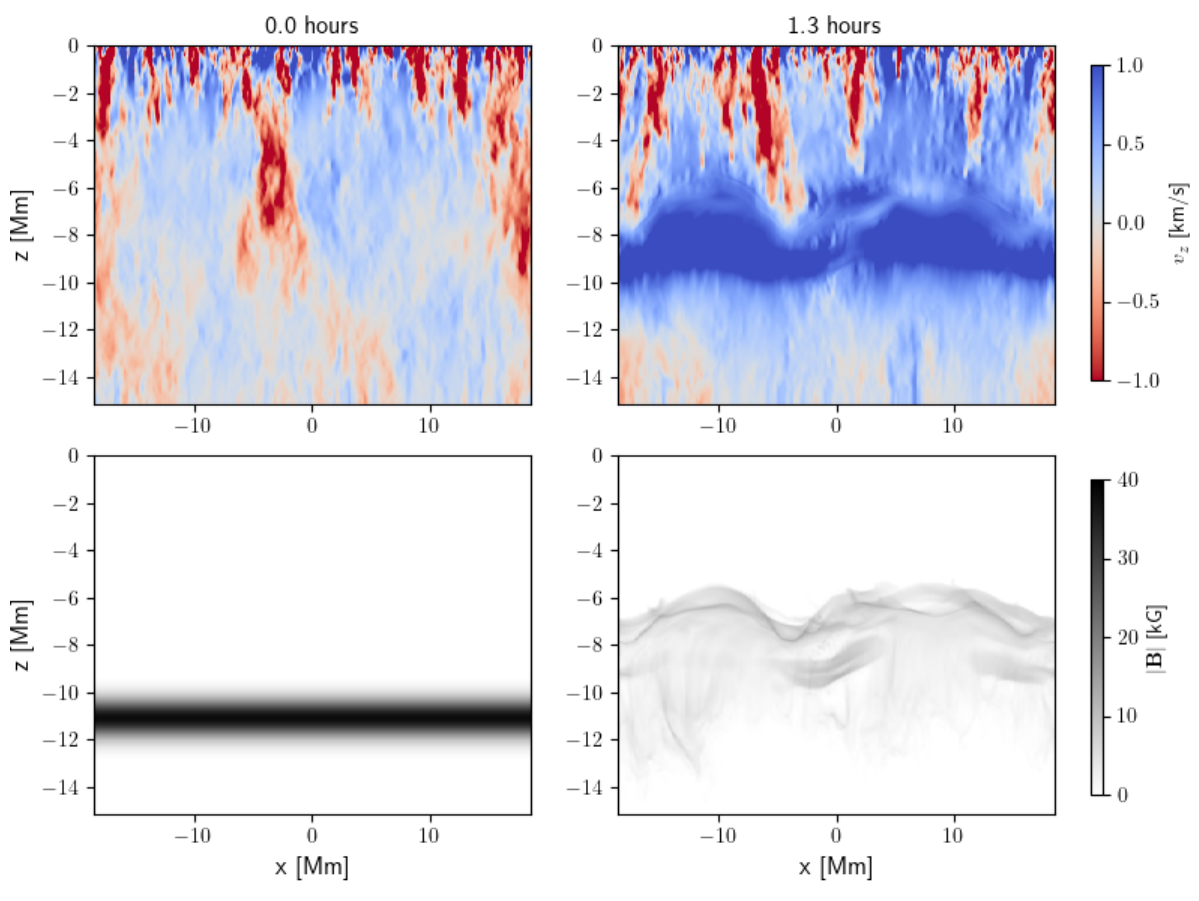}
    \vspace{-0.7cm}
    \caption{Vertical cuts through the middle of the $y=0$~Mm of the initial condition (left, $t=0$~hours) of a rising flux tube simulation with vertical velocity (top row, blue is upflow and red is downflow) and  absolute magnetic field strength (bottom row). 
    The right column shows the same at $t=1.3$~hours later, showing the effect of the large scale down flow on the rising flux tube. Note that the peak magnetic flux density of the tube of the initial condition (lower left panel) is 50~kG and the colour scale is saturated at 40~kG.}
    \label{fig:ictube}
\end{figure*}

We first computed a purely hydrodynamic initial condition which was run for enough time to relax and large scale structure to develop. 
We then placed a  flux tube centred at 11.096~Mm below the surface.
The flux tube lies in the $x$-direction with a purely horizontal magnetic field given in cylindrical coordinates centred on the tube axis ($r=0$ at $z=-11.096$~Mm) given as
\begin{equation*}
    B_x(r < 1.7~\mathrm{Mm},\theta,x)=B_0 e^{-X r^2},
\end{equation*}
 where $X=\pi B_0/\Phi_T$, $B_0 = 5 \times 10^4$~G is the maximum magnetic flux density, and $\Phi_T=10^{21}$~Mx is the total magnetic flux.
The FWHM of the radial profile is 1.3~Mm, and we set the magnetic field to zero beyond a radius of 1.7~Mm
so that the tube is localised at that depth (see Fig.~\ref{fig:ictube}). The magnetic field strength at 1.7~Mm is 760.9~G, or 1.52\% of the peak at the centre.

We constructed the tube to satisfy the solenoidal constraint, with all other components of the magnetic field equal to zero, and so that the magnetic flux is conserved along the length of the tube.  
To balance the pressure, we compensated the additional magnetic pressure within the flux tube by lowering the internal gas pressure, so that the total pressure is equal to the gas pressure before we added the tube. 
We kept the entropy unchanged from the hydrodynamic snapshot, and found the values for all thermodynamic quantities based on the entropy and gas pressure. This resulted in a lower density within the flux tube, with a peak density perturbation of less than 1\% along its axis.

We ran the simulation for a total of 7~hours, and a bipolar structure formed at about 5.4~hours (see discussion in Sect.~\ref{sect:sims} and Fig.~\ref{fig:bpflows}) with a lifetime of about 1~hour. 

\begin{figure}
	\includegraphics[width=0.5\textwidth]{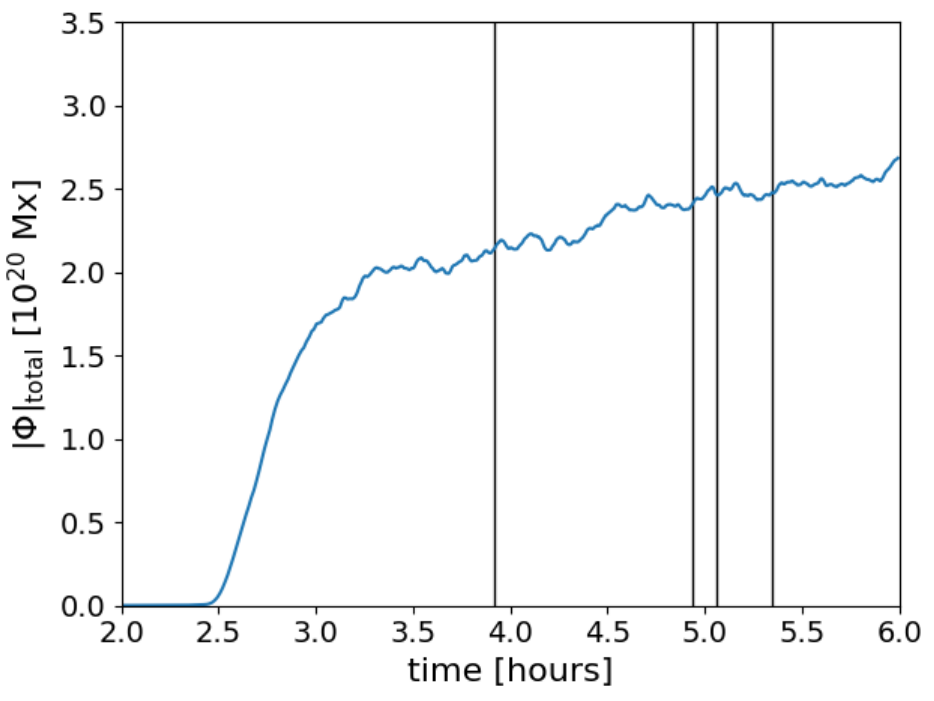}
    \vspace{-0.7cm}
    \caption{Total unsigned magnetic flux at the surface of the simulation. The vertical lines correspond to the times of the frames shown in Fig.~\ref{fig:bpflows}. }
    \label{fig:totphi}
\end{figure}

\bsp	
\label{lastpage}
\end{document}